\documentclass[usegraphicx,usenatbib]{mn2e}

\newcommand{\subs}[1]{$_{\rm #1}$}
\newcommand{\sups}[1]{$^{\rm #1}$}

\newcommand{\BE}{\begin{equation}}
\newcommand{\EE}{\end{equation}}

\newcommand{\kmss}{km\ s$^{-1}$ }
\newcommand{\vsinis}{$\!${\em v\,}sin{\em i} }
\newcommand{\vsini}{$\!${\em v\,}sin{\em i}}
\def\ga{\mathrel{\hbox{\rlap{\hbox{\lower4pt\hbox{$\sim$}}}\hbox{$>$}}}}
\def\la{\mathrel{\hbox{\rlap{\hbox{\lower4pt\hbox{$\sim$}}}\hbox{$<$}}}}

\title[The differential rotation of HD 141943]{Magnetic fields and differential rotation on the pre-main sequence II: The early-G star HD 141943 - coronal magnetic field, H$\alpha$ emission and differential rotation}
\author[S. C. Marsden et al.]
  {S.~C.~Marsden,$^{1,2}$\thanks{Email: Stephen.Marsden@jcu.edu.au (SCM); mmj@st-and.ac.uk (MMJ); jramirez@astroscu.unam.mx (JCRV); evelyne.alecian@obs.ujf-grenoble.fr (EA); brownca@usq.edu.au (CJB); carterb@usq.edu.au (BDC); donati@obs-mip.fr (J-FD); n.j.dunstone@googlemail.com (ND); rhodes.hart@usq.edu.au (RH); Meir.Semel@obspm.fr (MS); waite@usq.edu.au (IAW)} M.~M.~Jardine,$^3$\footnotemark[1] J.~C.~Ram\'{i}rez V\'{e}lez,$^{4,5}$\footnotemark[1]  E.~Alecian,$^{4,6}$\footnotemark[1] \and C.~J.~Brown,$^7$\footnotemark[1] B.~D.~Carter,$^7$\footnotemark[1] J.-F.~Donati,$^8$\footnotemark[1] N.~Dunstone,$^3$\footnotemark[1] R.~Hart,$^7$\footnotemark[1] \and M.~Semel$^4$\footnotemark[1]  and I.~A.~Waite$^7$\footnotemark[1]\\
   $^1$Centre for Astronomy, School of Engineering and Physical Sciences, James Cook University, Townsville, 4811, Australia\\
   $^2$Australian Astronomical Observatory, PO Box 296, Epping NSW 1710, Australia\\
   $^3$SUPA, School of Physics and Astronomy, University of St. Andrews, St. Andrews, KY 16 9SS, Scotland\\ 
   $^4$LESIA, Observatoire de Paris-Meudon, F-92195 Meudon Cedex, France\\
   $^5$Instituto de Astronomia, Universidad Nacional Aut\'{o}noma de M\'{e}xico, 04510, Coyoacon, M\'{e}xico D.F.\\
   $^6$LAOG, Laboratoire d'Astrophysique de Grenoble, Universit\'{e} Joseph Fourier, BP 53 38041, Grenoble Cedex 09, France\\
   $^7$Faculty of Sciences, University of Southern Queensland, Toowoomba, 4350, Australia\\
   $^8$LATT--UMR 5572, CNRS \& Univ.\ de Toulouse, 14 Av.\ E.~Belin, F--31400 Toulouse, France}
\date{Accepted version}

\pagerange{\pageref{firstpage}--\pageref{lastpage}} \pubyear{2010}

\begin{document}

\label{firstpage}

\maketitle

\begin{abstract}
Spectropolarimetric observations of the pre-main sequence early-G star HD 141943 were obtained at three observing epochs (2007, 2009 and 2010). The observations were obtained using the 3.9-m Anglo-Australian telescope with the UCLES echelle spectrograph and the SEMPOL spectropolarimeter visitor instrument. The brightness and surface magnetic field topologies (given in Paper I) were used to determine the star's surface differential rotation and reconstruct the coronal magnetic field of the star.

The coronal magnetic field at the 3 epochs shows on the largest scales that the field structure is dominated by the dipole component with possible evidence for the tilt of the dipole axis shifting between observations. We find very high levels of differential rotation on HD 141943 ($\sim$8 times the solar value for the magnetic features and $\sim$5 times solar for the brightness features) similar to that evidenced by another young early-G star, HD 171488. These results indicate that a significant increase in the level of differential rotation occurs for young stars around a spectral type of early-G. Also we find for the 2010 observations that there is a large difference in the differential rotation measured from the brightness and magnetic features, similar to that seen on early-K stars, but with the difference being much larger. We find only tentative evidence for temporal evolution in the differential rotation of HD 141943. 
\end{abstract}


\section{Introduction} \label{Sec_int}

One of the key drivers of the solar dynamo is differential rotation. In the Sun strong shears occur in the interface layer between the differentially rotating outer convective zone and the inner radiative zone, which rotates as a solid body. It is these shears that help convert the large-scale solar poloidal field into a strong toroidal component. However, for young, rapidly-rotating, solar-type stars a fundamentally different dynamo may be in operation.

Spectropolarimetric observations of young solar-type stars \citep[i.e.][]{DonatiJF:1997a, DonatiJF:1999a, DonatiJF:1999b, DonatiJF:2003a, MarsdenSC:2006, DunstoneNJ:2008, JeffersSV:2008} have shown that their reconstructed magnetic topologies have large regions of near-surface azimuthal field. These are interpreted as the toroidal component of the large-scale dynamo field in the stars. The presence of these regions near the stellar surface has led to the belief that the dynamo operating in such stars is in fact distributed throughout the stellar convective zone, rather than being restricted to the interface layer as in the solar case. 

Differential rotation is still thought to play a role in the generation of magnetic fields in these stars, however how the dynamos in these stars operates is not well understood. Most theoretical models are based on our knowledge of the solar dynamo. The models of \citet{KitchatinovLL:1999} predict that the level of differential rotation on an early-G star should be greater than that on a mid-K star and that the level of surface differential rotation should decrease for stars with shorter rotational periods. While the models of \citet{KukerM:2005} also show that the level of differential rotation on a star should be dependent upon its effective temperature (with hotter stars having higher levels of differential rotation) but only weakly dependent on its rotation rate. These are predictions that we can now observationally test.

There are several techniques we can use to observationally measure the level of differential rotation on a star. Direct starspot tracking from multiple Doppler images of the surface features \citep*[i.e.][]{CameronAC:2002}, cross-correlating two independent Doppler images \citep[i.e.][]{DonatiJF:1997a}, incorporating a differential rotation law into the Doppler imaging process \citep*[i.e.][]{DonatiJF:2000,PetitP:2002}, or by Fourier analysis of stellar line profiles \citep[i.e.][]{ReinersA:2002, ReinersA:2003}.

Combining differential rotation measurements of a number of young solar-type stars, found using the Doppler imaging method, \citet{BarnesJR:2005} found that the level of differential rotation increases with stellar temperature with early-G stars having higher levels of differential rotation than lower-mass stars, in agreement with the findings of \citet{KitchatinovLL:1999} and \citet{KukerM:2005}. The results also showed only a weak (if any) correlation with stellar rotation rate, again in agreement with the findings of \citet{KukerM:2005}. However, more recent measurements of the young early-G star HD 171488 \citep{MarsdenSC:2006, JeffersSV:2008, JeffersSV:2010} have shown this star to have an even higher level of differential rotation than that indicated by \citet{BarnesJR:2005} with differential rotation measurements up to 10 times the current solar value. This supports the findings of \citet{ReinersA:2006} which show high levels of differential rotation on a number of F stars. 

Work by \citet*{DonatiJF:2003b} has shown that for some early-K stars the level of differential rotation measured from surface brightness features is consistently lower than that measured from the magnetic features (using the same dataset). This has led them to surmise that this is a result of the magnetic and brightness features being anchored at different depths within the stellar convective zone and that the convective zone has a radially varying differential rotation structure, unlike the Sun. The work also showed that these early-K stars evidence temporal variation in their levels of differential rotation that the authors attribute to a feed-back mechanism in the stellar dynamo periodically converting magnetic energy into kinetic energy and vice-versa.

As mentioned there is currently only one early-G star for which there are differential rotation measures from both brightness and magnetic features obtained at multiple epochs, HD 171488 \citep{MarsdenSC:2006, JeffersSV:2008, JeffersSV:2010}. This work has shown very little difference in the level of differential rotation measured from the magnetic and brightness features, although the measurement errors are larger than that found on the early-K stars \citep{DonatiJF:2003b}. Additionally, there appears to be little evidence of temporal evolution in the level of differential rotation on HD 171488. This has been speculated to be caused by the thinner convective zone of HD 171488 compared to that of the early-K stars previously studied.

In order to expand the number of young early-G stars studied with spectropolarimetry this paper along with the first paper in the series \citet[][Paper I]{MarsdenSC:2010} and \citet{WaiteIA:2010}, presents differential rotation measurements, and magnetic maps, of two young early-G pre-main sequence (PMS) stars. Paper I deals with the reconstruction of the brightness and magnetic topologies of the star HD 141943. This paper (Paper II) presents the coronal magnetic field reconstructions, H$\alpha$ activity and differential rotation measurements of HD 141943. The third paper in the series \citep{WaiteIA:2010} deals with observations of HD 106506.

As detailed in Paper I, HD 141943 is a young, active and rapidly-rotating PMS star. The stellar parameters have been determined in Paper I through the Doppler imaging process and are reproduced here in Table~\ref{Tab_par}.

\begin{table}
\caption{Fundamental parameters of HD 141943 from Paper I.}
\label{Tab_par}
\centering
\begin{tabular}{lc}
\hline\hline
Parameter & value\\
\hline
Age & $\sim$17 Myrs\\
Mass & $\sim$1.3 M\subs{\odot}\\
Photospheric temperature & 5850 $\pm$ 100 K$^{a}$\\
Spot temperature & $\sim$3950 K\\
Unspotted luminosity & 2.8 $\pm$ 0.1 L$^{a}_{\odot}$\\
Stellar radius & 1.6 $\pm$ 0.15 R\subs{\odot}\\
\vsinis & 35.0 $\pm$ 0.5 \kmss\\
Radial velocity ($v_{\rm rad}$) & $\sim$0.1 \kmss\\
Inclination angle ($i$) & 70\sups{\circ} $\pm$ 10\sups{\circ}\\
Equatorial rotation period ($P_{\rm eq}$) & $\sim$2.182 days\\
\hline
$^{a}$assumed errors (see Paper I).
\end{tabular}
\end{table}

The reconstructed brightness images show that it has a weak polar spot and a significant amount of low-latitude spot features at all epochs. The magnetic reconstructions show that its has a predominately non-axisymmetric radial field while its azimuthal field is predominately axisymmetric with ring of azimuthal field seen at the pole, similar to that of other active stars.

\section{Observations and data reduction} \label{Sec_obs}

HD141943 was observed at 3 epochs on the 3.9-m Anglo-Australian telescope (AAT), in March/April 2007, April 2009 and March/April 2010. A fourth observation set was taken in May 2006 (see Paper I) but due to the limited nature of this set and the fact that it was only spectroscopic observations we have not included it in our analysis here (Note: a differential rotation measure was attempted on this dataset but it was too limited to provide a result).The three datasets used were spectropolarimetric observations in left- and right-hand circularly polarised light taken using the University College London Echelle Spectrograph (UCLES) and the SEMPOL spectropolarimeter \citep*{SemelM:1993, DonatiJF:2003a} visitor instrument. 

The data were reduced using the ESpRIT (Echelle Spectra Reduction: an Interactive Tool) optimal extraction routines of \citet{DonatiJF:1997b}. We then used the technique of Least-Squares Deconvolution \citep[LSD,][]{DonatiJF:1997b} to sum the over 2600 photospheric spectral lines in each echelle spectrum in order to create a single high signal-to-noise (S/N) profile for each observation. Further details of the observations, including an observing log, and the reduction process are given in Paper I.

\section{Results} \label{Sec_res}

Brightness and magnetic maps of HD 141943 were created at the three epochs through the inversion of the observed Stokes I (brightness) and Stokes V (magnetic) LSD profiles. These images and the method used to create them are given in Paper I. As an extension to these results we have measured the differential rotation on the surface of HD 141943 using these maps. We have also extrapolated the radial magnetic field map to reconstruct the coronal magnetic field of the star and we have looked at the H$\alpha$ emission from the star to look for prominence activity.

\subsection{Surface differential rotation} \label{Sec_sdr}

If a star is observed for a number of rotations, then the surface features will evolve under the influence of the star's surface differential rotation. The surface differential rotation on a star can be determined from both its spot/brightness features as well as from its magnetic features. 

In order to measure the differential rotation on a star a simplified solar-like differential rotation law is incorporated into the imaging process:
\begin{equation}
\Omega(\theta) = \Omega_{\rm eq} - d\Omega sin^{2}\theta {\rm (rad\ d^{-1})},
\end{equation}
where $\Omega(\theta)$ is the rotation rate at latitude $\theta$, $\Omega$\subs{eq} is the equatorial rotation rate and $d\Omega$ is the rotational shear between the equator and the poles (the differential rotation). The surface differential rotation is then determined by treating both $\Omega$\subs{eq} and $d\Omega$ as free parameters and determining the best fit to the data using the $\chi^{2}$-minimisation method. This method is further described in \citet{DonatiJF:2000, DonatiJF:2003b} and \citet{PetitP:2002}.

To determine the surface differential rotation usually requires that a significant part of the stellar surface is observed on at least two occasions preferably several days apart \citep[see][]{PetitP:2002}. As the rotational period of HD 141943 is close to 2.2 days (see Table~\ref{Tab_par}) this means that in order to determine the level of surface differential rotation the observations need to cover a time base of around a week or more. We attempted to determine the level of surface differential rotation on HD 141943 from the brightness and magnetic features in each dataset. Unfortunately, the differential rotation could only be measured from the March/April 2007 magnetic data and the March/April 2010 magnetic and brightness data. The other datasets did not produce a minimum in the $\chi^{2}$-landscape (see below) so the level of differential rotation could not be determined.

\begin{figure}
  \centering
  \includegraphics[angle=-90, width=\columnwidth]{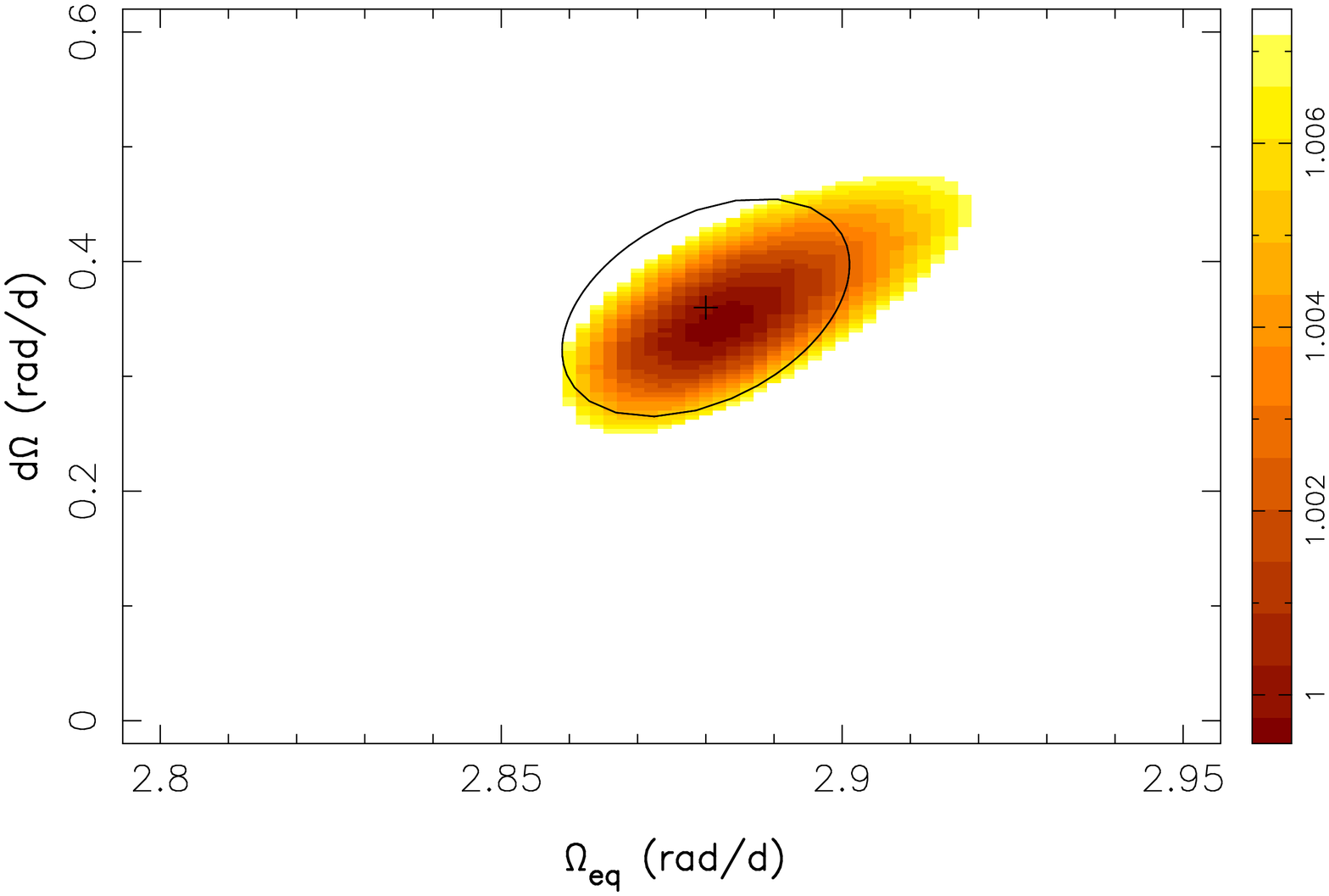}
  \includegraphics[angle=-90, width=\columnwidth]{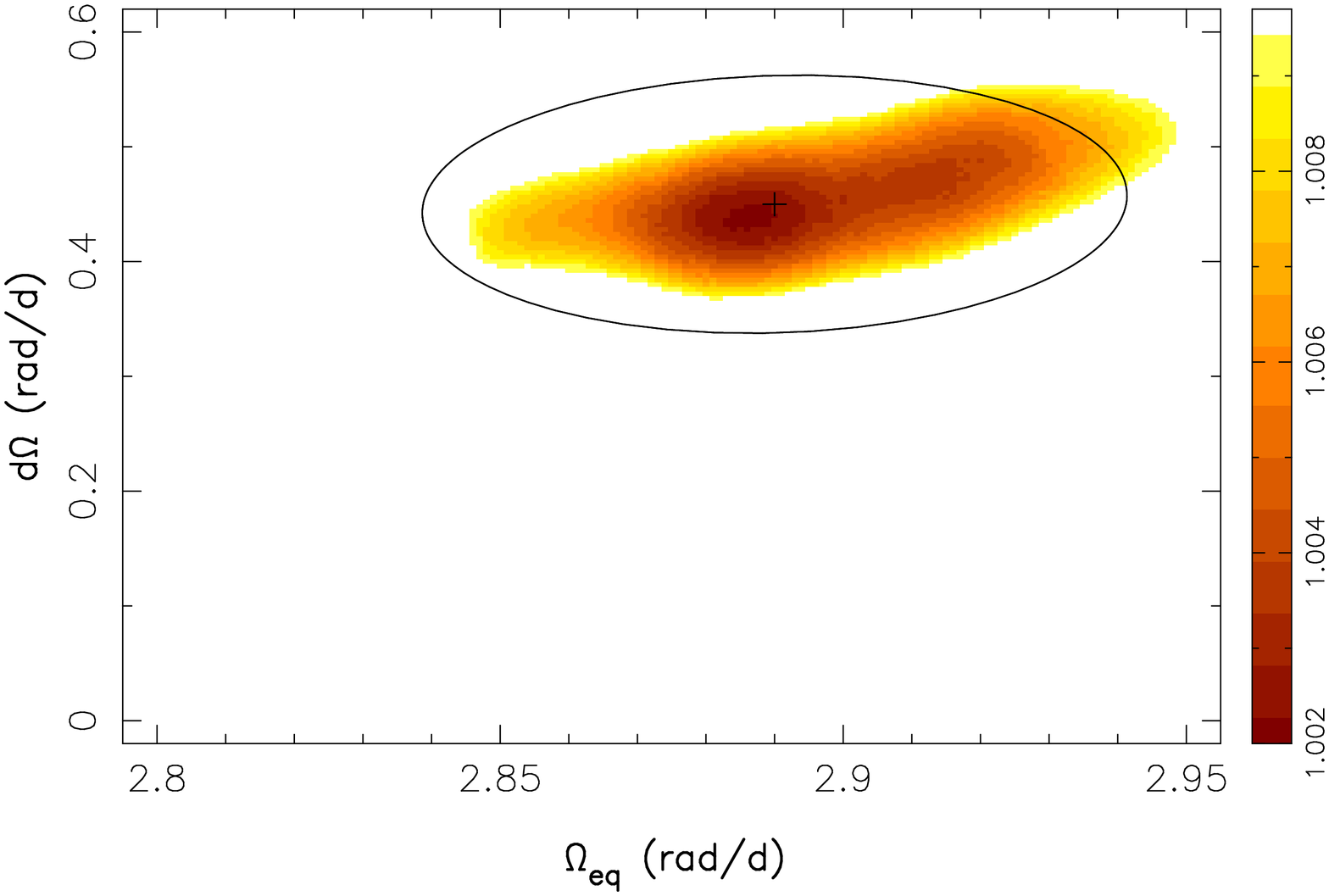}
  \includegraphics[angle=-90, width=\columnwidth]{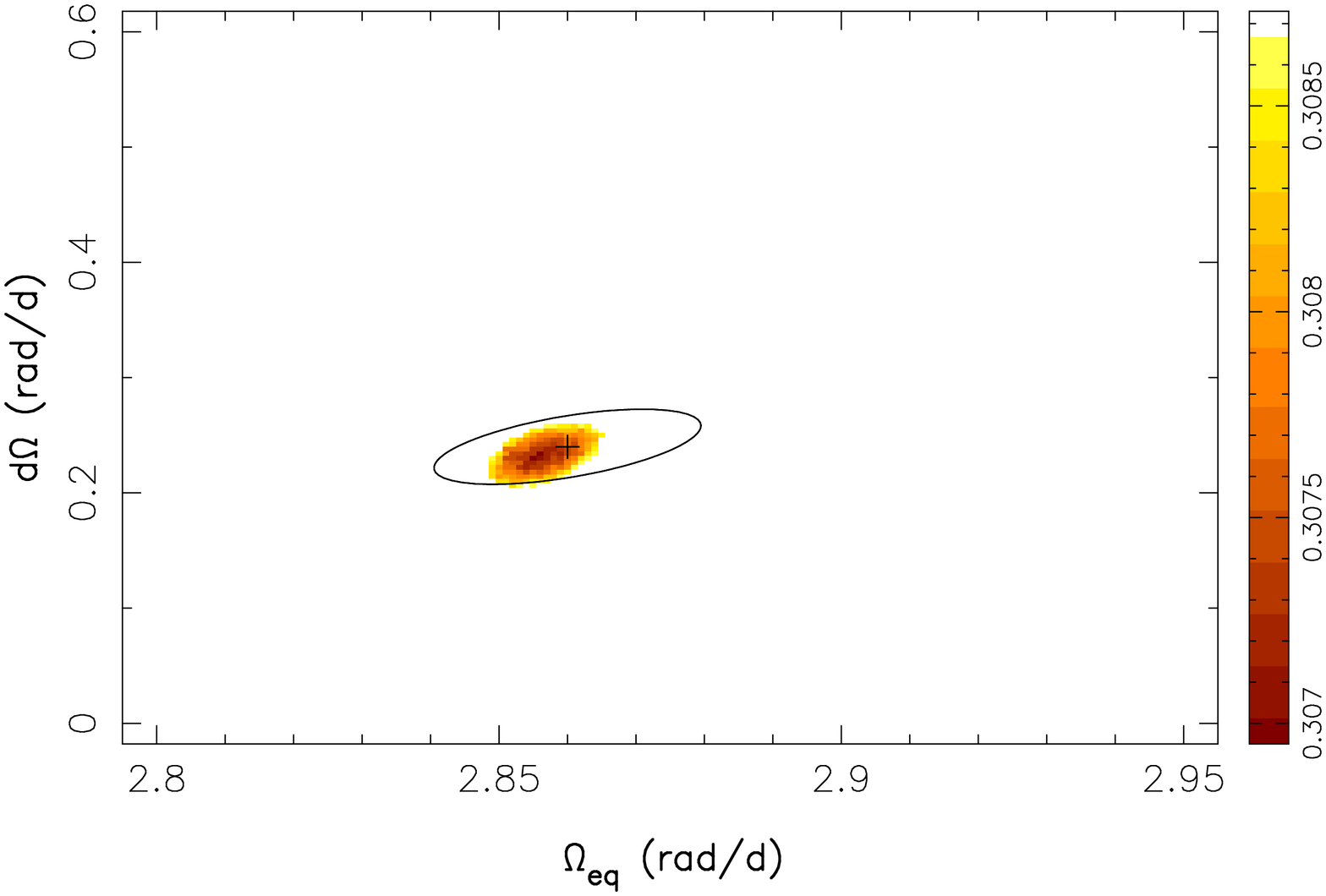}
  \caption{Surface differential rotation $\chi^{2}$-minimisation for HD141943, March/April 2007 Stokes V data (upper plot), March/April 2010 Stokes V data (middle plot) and March/April 2010 Stokes I data (bottom plot). The coloured images show the reduced $\chi^{2}$ values obtained from the maximum-entropy imaging code for various values of $\Omega$\subs{eq} and $d\Omega$ assuming all stellar parameters are correct. Darker regions correspond to lower reduced $\chi^{2}$ values. The coloured images project to $\pm$3$\sigma$ on both axes for the upper and middle plots and $\pm$5$\sigma$ for the lower plot.  The overplotted ellipses show the 1$\sigma$ errors (projected onto both axes) for the differential rotation measures when taking into account the errors in the stellar parameters.}
  \label{Fig_dr}
\end{figure}

For the March/April 2007 magnetic data (Stokes V) the imaging code was forced to converge to a fixed magnetic field value of 91.3 G (with the level of magnetic field determined through an iterative process) for various values of $\Omega$\subs{eq} and $d\Omega$. This produced the reduced $\chi^{2}$-landscape shown in the upper image of Fig.~\ref{Fig_dr}. Fitting a paraboloid to this data gave $\Omega$\subs{eq} = 2.882 $\pm$ 0.009 rad d\sups{-1} and $d\Omega$ = 0.347 $\pm$ 0.035 rad d\sups{-1}, with the errors being 1$\sigma$ errors.

For the March/April 2010 magnetic data (Stokes V) the imaging code was forced to converge to a fixed magnetic field value of 70.9 G and produced the reduced $\chi^{2}$-landscape shown in the middle image of Fig.~\ref{Fig_dr}. Fitting a paraboloid to the central region of the data gave $\Omega$\subs{eq} = 2.886 $\pm$ 0.009 rad d\sups{-1} and $d\Omega$ = 0.439 $\pm$ 0.025 rad d\sups{-1}.

The level of differential rotation ($d\Omega$) in 2010 for the magnetic features is higher that that found in 2007, with the difference corresponding to a $\sim$2$\sigma$ change. This is a marginal change so we feel that it is only tentative evidence of temporal evolution in the differential rotation of HD 141943.

For the March/April 2010 brightness data (Stokes I) the imaging code was forced to converge to a spot filling factor of 0.029 (2.9 per cent) and produced the reduced $\chi^{2}$-landscape shown in the bottom image of Fig.~\ref{Fig_dr}. Fitting a paraboloid gave: $\Omega$\subs{eq} = 2.856 $\pm$ 0.002 rad d\sups{-1} and $d\Omega$ = 0.232 $\pm$ 0.007 rad d\sups{-1}.

\begin{figure}
  \centering
  \includegraphics[angle=-90, width=\columnwidth]{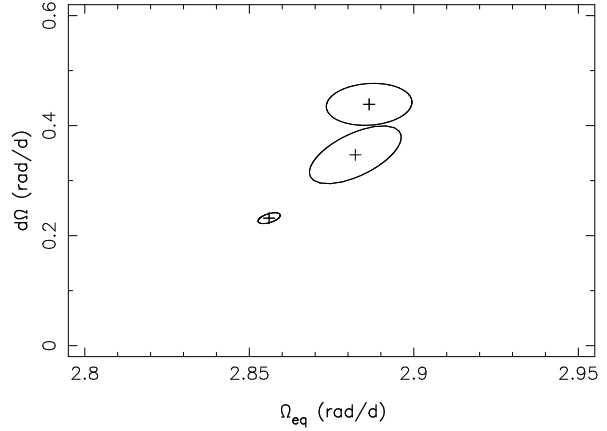}
  \caption{Plot showing the 1$\sigma$ error ellipses for the surface differential rotation for HD 141943. The upper ellipse is the Stokes V data from 2010, the middle ellipse is the Stoves V data from 2007 and the bottom ellipse is the Stokes I data from 2010. These error ellipses have not taken into account the errors in the stellar parameters (see text).} 
  \label{Fig_ee}
\end{figure}

For 2010 the level of differential rotation from the brightness features is significantly different to that found from the magnetic features, corresponding to a $\sim$8$\sigma$ change. This is a definite detection of different differential rotation rates for the brightness and magnetic features on HD 141943. Fig.~\ref{Fig_ee} shows a plot of the 1$\sigma$ error ellipses for all three differential rotation measurements on the same graph.

These measures (and errors) of differential rotation assume that the stellar parameters used in the imaging code are correct and as shown by Table~\ref{Tab_par} they themselves have some uncertainty. The main parameters that will affect the measure of differential rotation are; the \vsini, inclination angle and rotational period. Taking these errors into account gave the surface differential rotation measures shown in Table~\ref{Tab_dr} and also as the overplotted ellipses in Fig.~\ref{Fig_dr}. 

\begin{table}
\caption{Surface differential rotation measurements for HD 141943, taking into account the errors in the stellar parameters given in Table~\ref{Tab_par}.}
\label{Tab_dr}
\centering
\begin{tabular}{ccc}
\hline\hline
Epoch      & $\Omega$\subs{eq} & $d\Omega$        \\
                  & (rad d\sups{-1})         & (rad d\sups{-1}) \\ 
\hline
                  & Stokes I (brightness) &                              \\
2010.244 & 2.86 $\pm$ 0.02        & 0.24 $\pm$ 0.03 \\
\\
                  & Stokes V (magnetic) &                                \\
2007.257 & 2.88 $\pm$ 0.02        & 0.36 $\pm$ 0.09  \\
2010.244 & 2.89 $\pm$ 0.05        & 0.45 $\pm$ 0.08  \\                  
\hline
\end{tabular}
\end{table}

The level of differential rotation for the magnetic features on HD 141943 is approximately 8 times the solar differential rotation rate. While the level of differential rotation for the brightness features is approximately 5 times the solar value.

It should be noted that the magnetic data for April 2009 gave an extremely shallow paraboloid around $\Omega$\subs{eq} = 2.84 $\pm$ 0.02 rad d\sups{-1} and $d\Omega$ = 0.0 $\pm$ 0.1 rad d\sups{-1}. However, the limited number of overlapping profiles and the poor quality of the observations early in the observing run (see Paper I) means that we believe this measurement of surface differential rotation to be highly suspect.

\subsection{Coronal field extrapolation}\label{Sec_cor}

The coronal magnetic fields are extrapolated from the surface magnetograms using the ``Potential Field Source Surface'' method \citep{AltschulerMD:1969,vanBallegooijenA:1998}. Since the method has been described in \citet*{JardineM:2002a} we provide only an outline here. Briefly, the condition that the field is potential requires that ($\vec{\nabla}\times\vec{B} =0$) and so we can write the magnetic field $\vec{B}$ in terms of a scalar flux function $\Psi$ such that $\vec{B} = -\vec{\nabla} \Psi$. The condition that the field is divergence-free then reduces to Laplace's equation $\vec{\nabla}^2 \Psi=0$ with a solution in spherical co-ordinates $(r,\theta,\phi)$
\begin{equation}
\Psi = \sum_{l=1}^{N}\sum_{m=-l}^{l} [a_{lm}r^l + b_{lm}r^{-(l+1)}] P_{lm}(\theta) e^{i m \phi},
\end{equation}
where the associated Legendre functions are denoted by $P_{lm}$. The coefficients $a_{lm}$ and $b_{lm}$ are determined by imposing the radial field at the surface from the Zeeman-Doppler maps and by assuming that at some height $R_s$ above the surface (known as the {\em source surface}) the pressure of the hot coronal gas overcomes the ability of the magnetic field to confine it. Thus, at the source surface the field lines are opened up to become purely radial, and hence  $B_\theta (R_s) = B_\phi(R_s) = 0$.

\begin{figure*}
 \begin{center}
  \includegraphics[width=7cm]{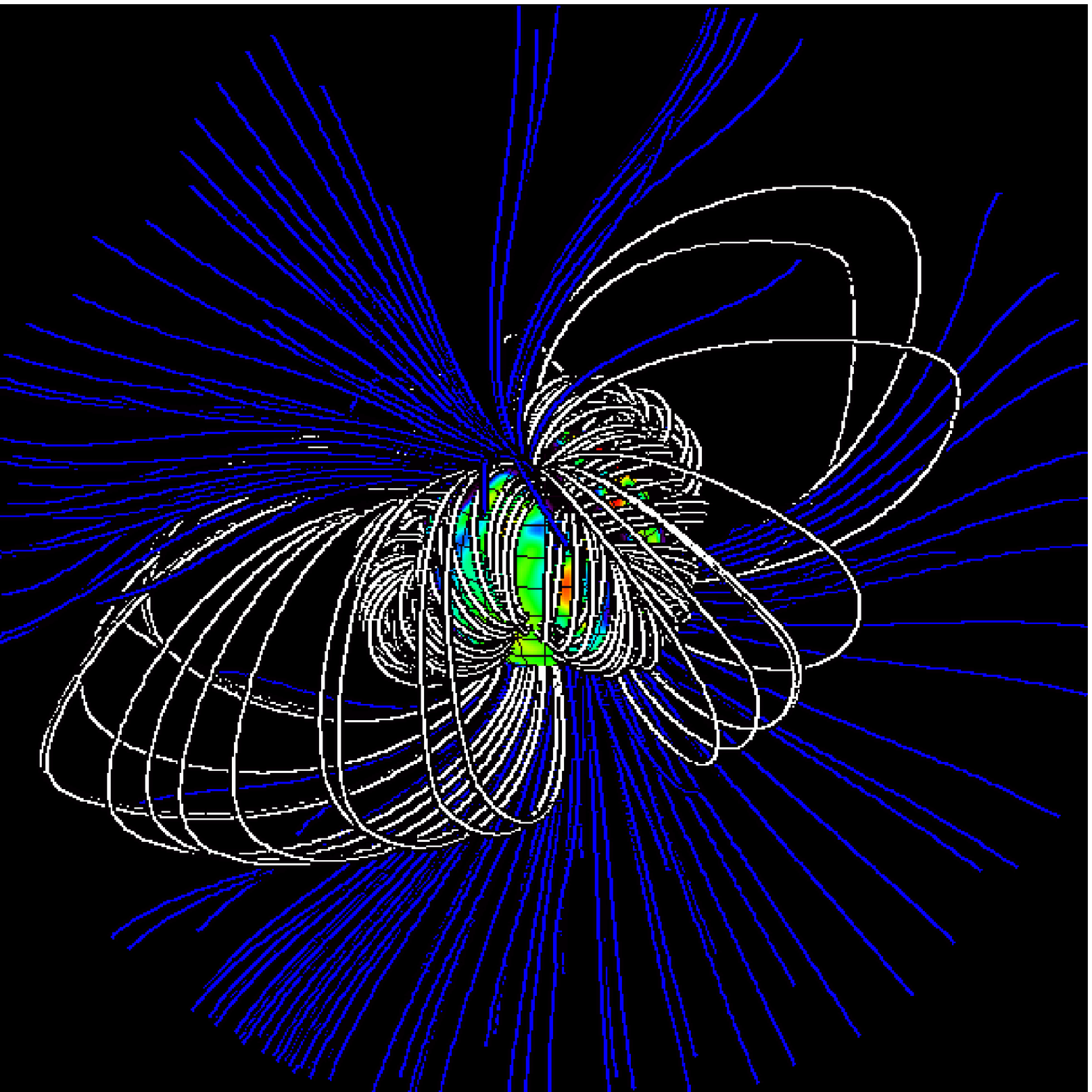}
  \includegraphics[width=7cm]{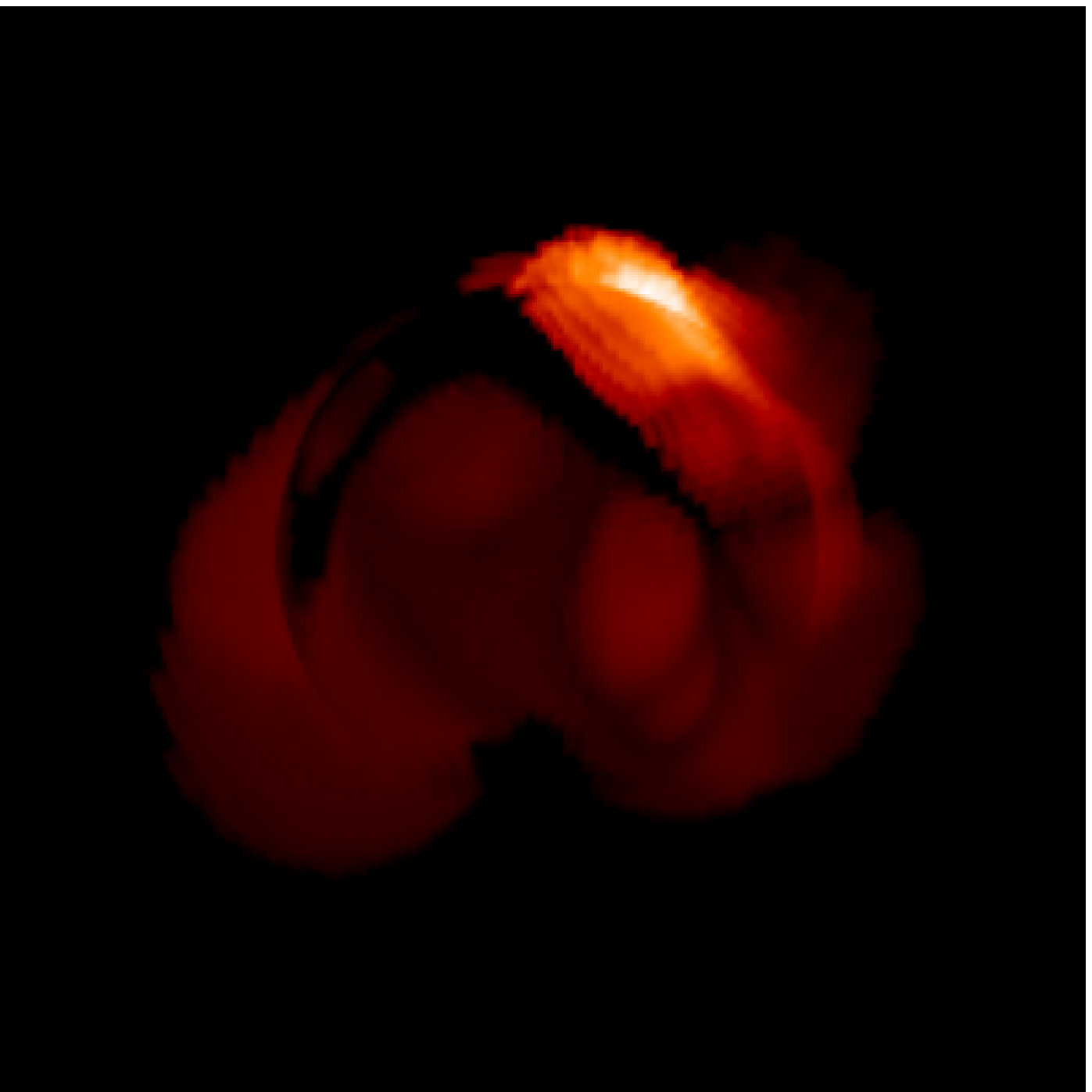}
  \includegraphics[width=7cm]{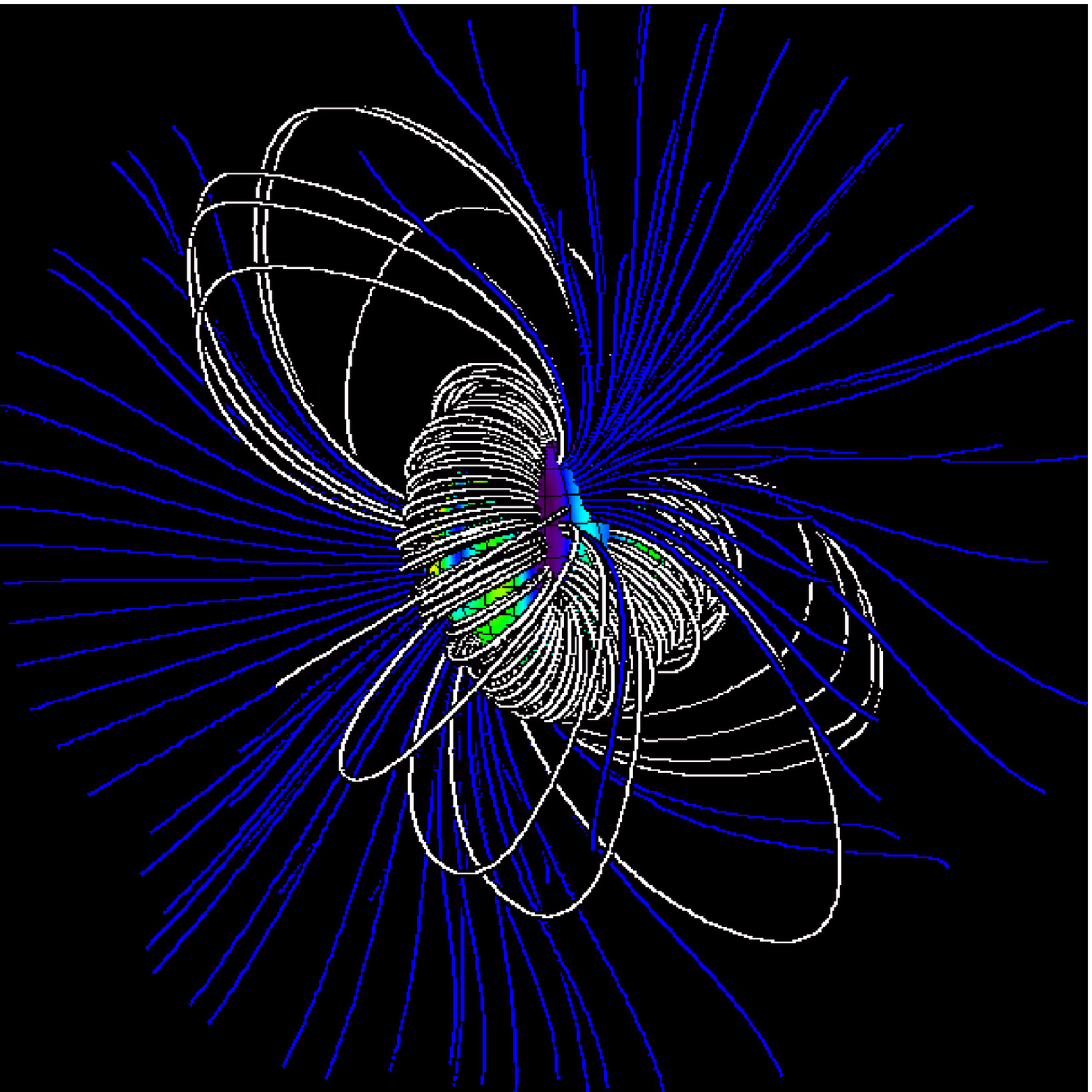}
  \includegraphics[width=7cm]{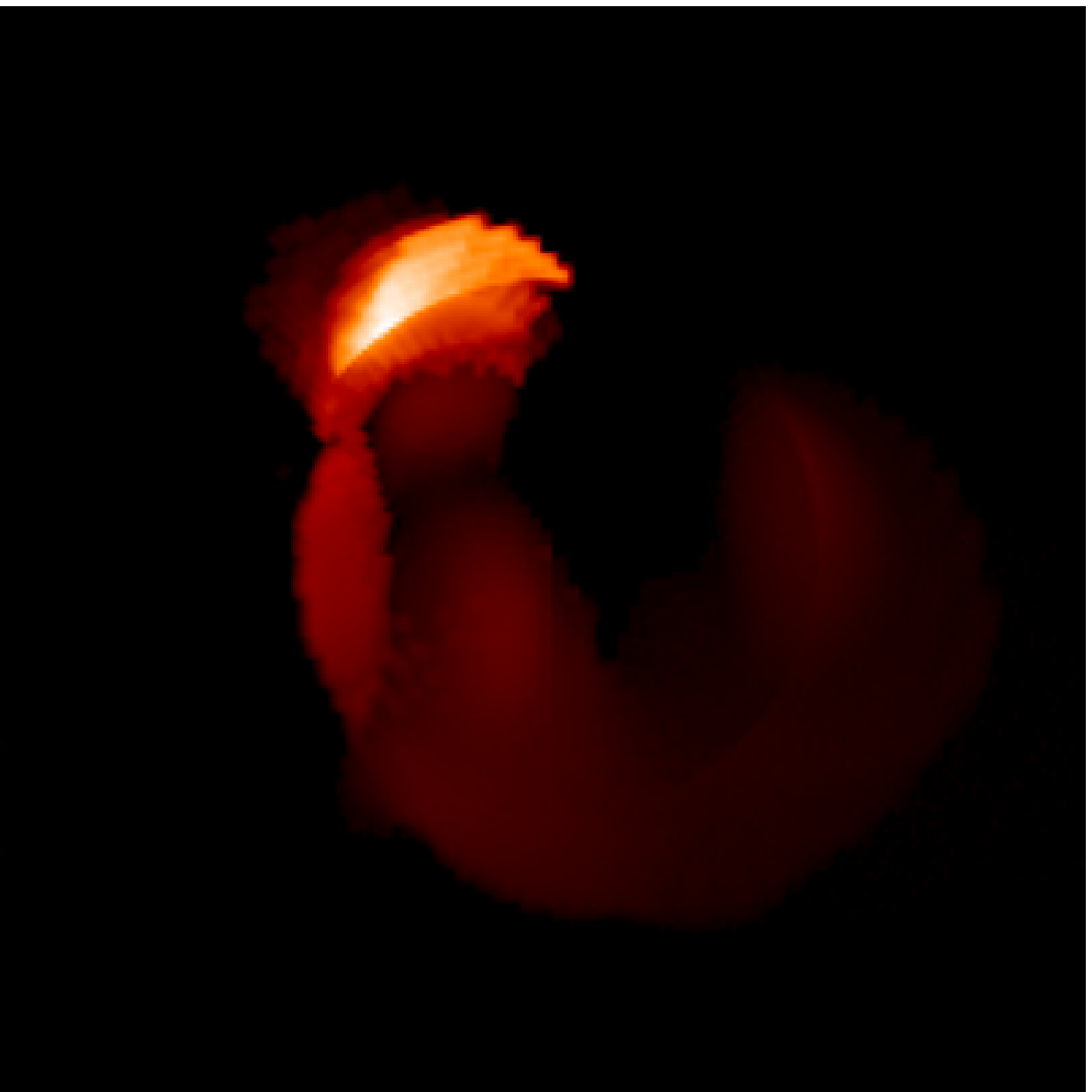}
  \includegraphics[width=7cm]{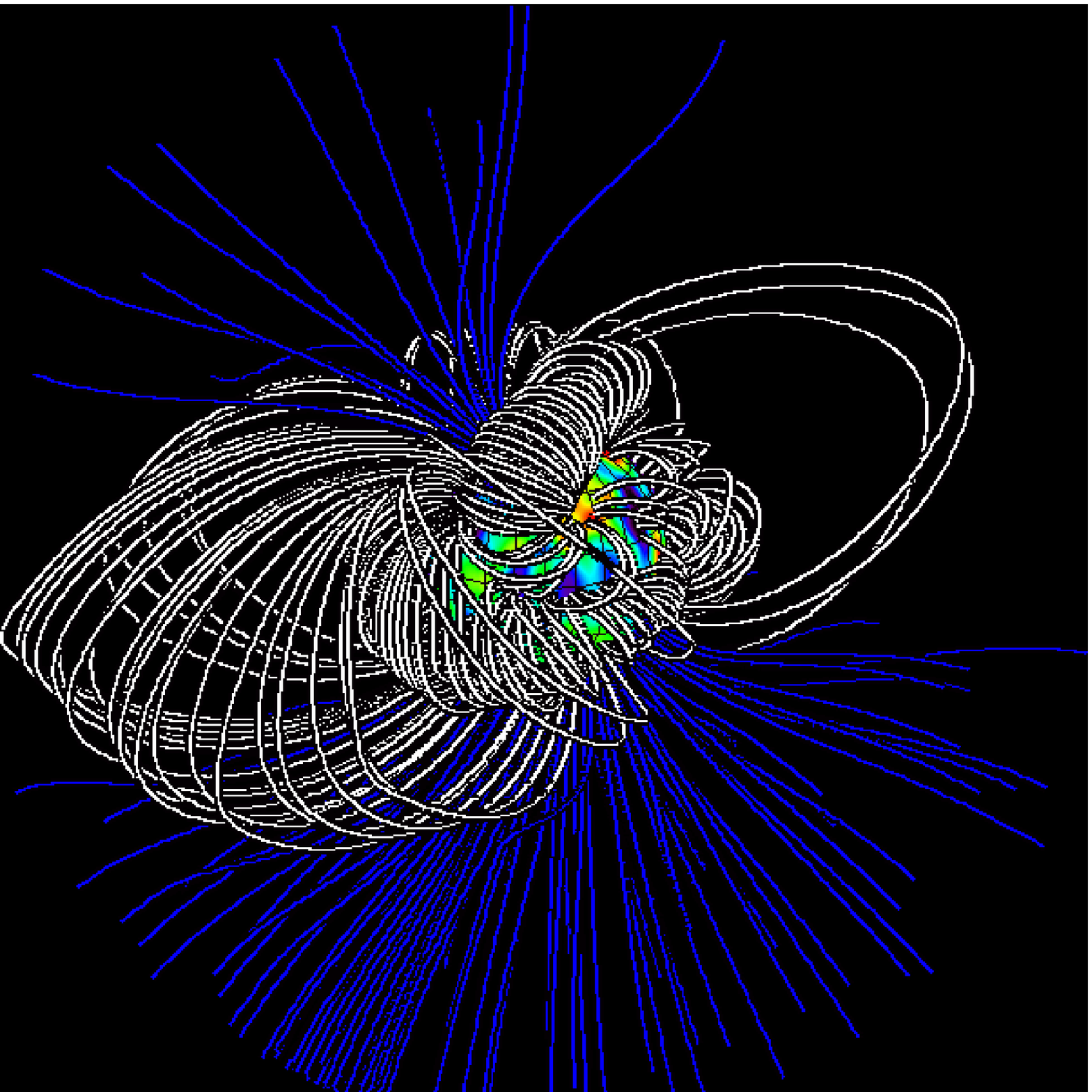}
  \includegraphics[width=7cm]{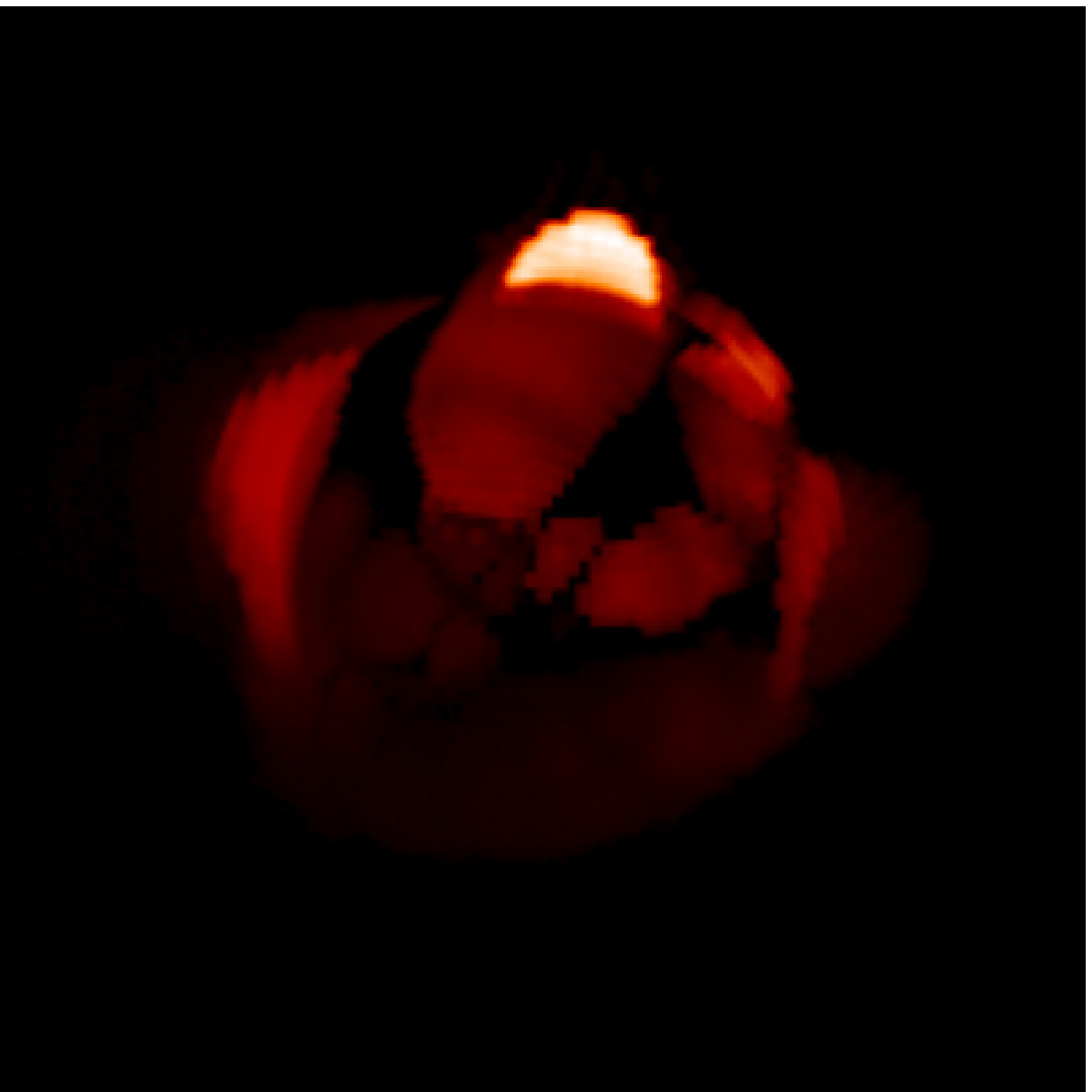}
   \caption{The images on the left are the coronal magnetic structure of HD 141943 in 2007 (upper-left), 2009 (middle-left) and 2010 (bottom-left) for a source surface of 4.8 R\subs{\star}. Closed field lines are shown in white while open field lines are shown in blue. The images on the right are the X-ray emission of HD 141943 in 2007 (upper-right), 2009 (middle-right) and 2010 (lower-right) for the same source surface and a coronal temperature of 2$\times$10\sups{7} K. All images are looking at phase 0.75 on the star.}
  \label{cor_struct}
  \end{center}
\end{figure*}

We determine the plasma pressure at every point by calculating the path of the field line through that point and solving for isothermal, hydrostatic equilibrium along that path. In this case, the gas pressure is $p=p_{0}e^{\frac{m}{k_B T}\int g_{s}ds}$ where $m$ is the mean particle mass, $k_B$ is Bolzmann's constant, T is the temperature and $p=p_{0}$ is the gas pressure at the base of the field line. We note that the integral in this expression is performed along the path of the field line and that $g_{s} =( {\bf g.B})/|{\bf B}|$ is the component of gravity along the field. We note that the plasma pressure is set to zero at any point where the field line through that point experiences a plasma pressure greater than the magnetic pressure somewhere along its length. In this case, we assume that this field line should have been forced open by the pressure of the plasma. The gas pressure at the footpoint of the field line $p_0$ is a free parameter of this model. Following \citet{JardineM:2002a,JardineM:2002b} we choose to scale $p_{0}$ to the magnetic pressure at the base of the field line, such that $p_{0}=K B^{2}_{0}$ where $K$ is a constant that is the same on every field line. By scaling $K$ up or down we can scale the overall level of the coronal gas pressure and hence the density and emission measure. Our model therefore has two parameters: the radius at which the field lines are opened up (the {\it source surface}) and the constant $K$ which determines the gas pressure $p_0$ at the base of each field line. These two parameters determine the magnetic field structure and  X-ray emission measure of the closed-field regions of the stellar corona. These are given in Fig.~\ref{cor_struct} for the three epochs.

\subsection{H$\alpha$ variation} \label{Sec_Ha}

\begin{figure}
  \centering
  \includegraphics[angle=90, width=\columnwidth]{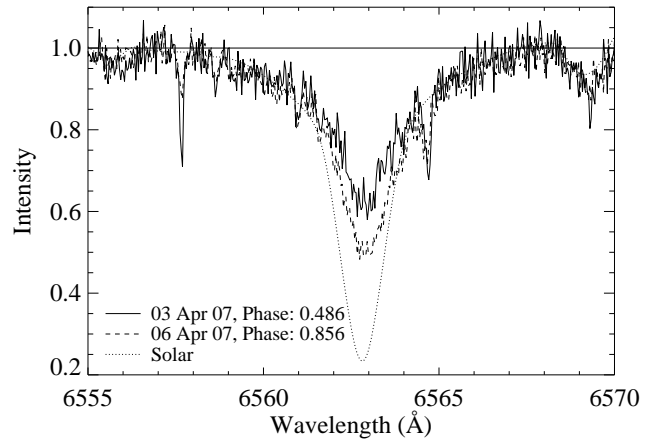}
  \caption{Plot of the H$\alpha$ profile of HD141943 at two epochs in March/April 2007 along with the solar H$\alpha$ profile taken with the same instrumental setup and shifted and convolved (by the \vsinis of HD 141943) to match the HD 141943 observations.}
  \label{Fig_ha}
\end{figure}

The H$\alpha$ line in active solar-type stars is often used as an activity indicator with the line being ``filled-in'' in more active stars \citep[i.e.][]{SoderblomDR:1993}. In addition, prominence activity in the stellar chromosphere can also be mapped on such stars that show emission in the H$\alpha$ line \citep[i.e.][]{CameronAC:1989, DonatiJF:2000}. We have analysed the H$\alpha$ line of HD 141943 in March/April 2007 and March/April 2010 as the most complete datasets to look for possible variations that may be attributable to prominence activity in the stellar chromosphere. As shown in Fig.~\ref{Fig_ha}, HD 141943 is a more active star than the Sun with a variable level of activity in 2007, but it does not have the overt H$\alpha$ emission of a young T Tauri star.

\begin{figure*}
  \centering
  \includegraphics[width=\columnwidth]{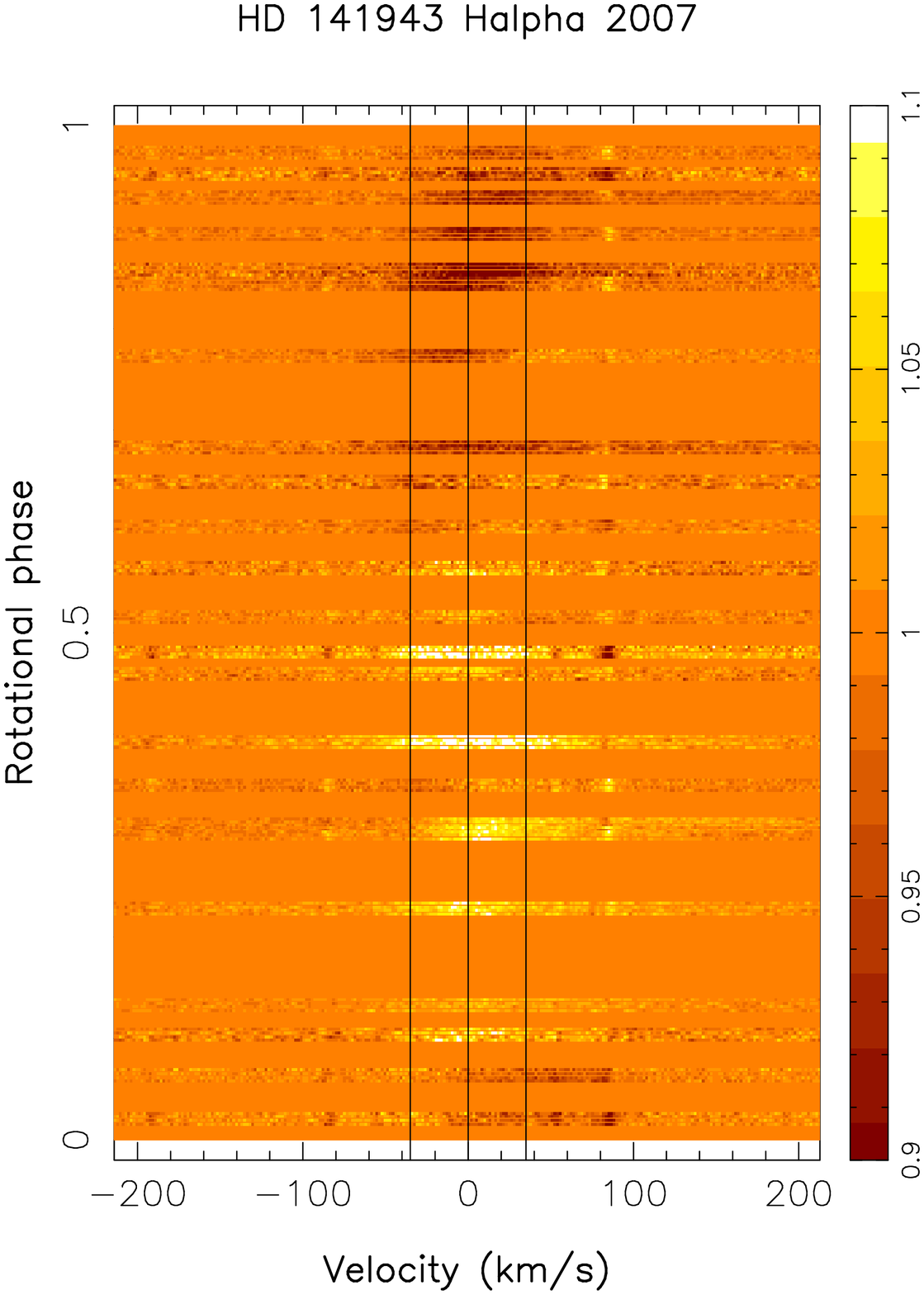}
  \includegraphics[width=\columnwidth]{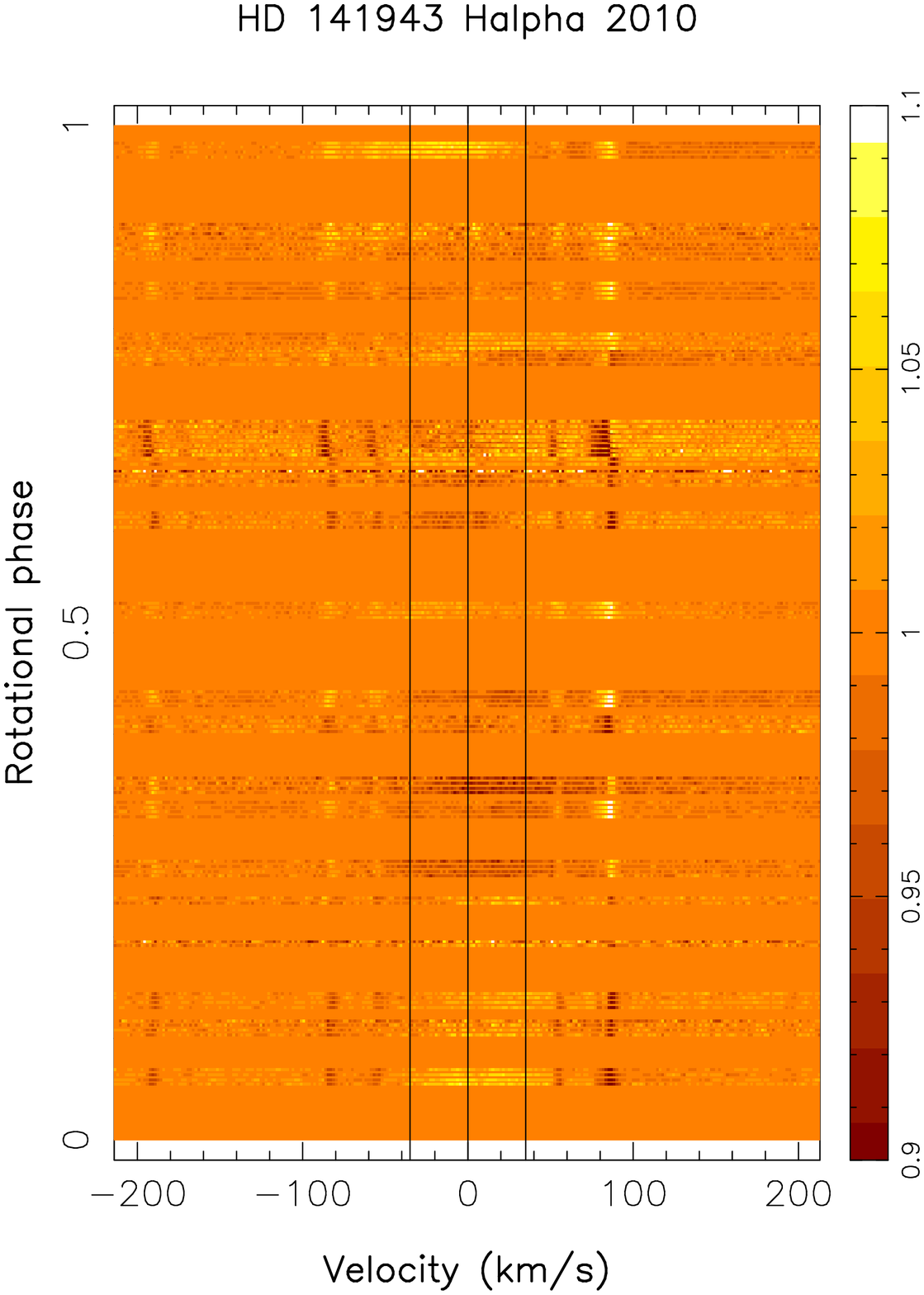}
  \caption{Dynamic spectrum of the variation in the H$\alpha$ profile of HD141943, March/April 2007 (left) and March/April 2010 (right).  All H$\alpha$ profiles in each dataset have been divided by the average H$\alpha$ profile of each dataset. The vertical thin lines represent the radial and rotational velocity of the star.}
  \label{Fig_dynha}
\end{figure*}

In order to look for prominence activity on HD 141943 we have divided each of the H$\alpha$ profiles in the March/April 2007 dataset by the mean H$\alpha$ profile from the dataset. We did the same for the March/April 2010 dataset, dividing by the average H$\alpha$ profile from the 2010 dataset. Dynamic spectra of these are displayed in Fig.~\ref{Fig_dynha} (2007 on the left and 2010 on the right) with darker areas showing regions of lower activity and lighter areas higher activity.

As can be seen the H$\alpha$ profile of HD 141943 in 2007 is variable in its activity level with regions of higher activity located between phases $\sim$0.1 to $\sim$0.6 while the other phases have lower activity than average. There would appear to be some evolution in the velocity of these active/non-active regions with most of the motion restricted to within the rotational velocity of the star, although there is some enhancement/reduction of the H$\alpha$ emission outside these velocities. Given the inclination angle of HD 141943, prominences located at the co-rotation radius of the star are likely to be seen as absorption features quickly crossing the \vsinis range of the star. The absorption features seen in Fig.~\ref{Fig_dynha} around phase $\sim$0.85 appear to travel across the stellar profile, but not rapidly. A sine wave fitted to the peak/troughs of the H$\alpha$ emission shows the amplitude of the sine wave to be restricted to within the \vsinis of HD 141943. Thus we believe that this region is located fairly close to the surface of the star and appears to be an inactive (in H$\alpha$) low-to-mid latitude feature on, or near, the stellar surface. Comparing this to the X-ray emission image (top-right image in Fig.~\ref{cor_struct}) we see that for phases around 0.0 there is little X-ray emission from the star.

For 2010 there appears to be little or no variation in the H$\alpha$ emission of HD 141943. The level of H$\alpha$ emission at all phases in 2010 is similar to the highest level of emission in 2007. Thus it would appear that the H$\alpha$ feature seen in 2007 is not present on the star in 2010.

\section{Discussion} \label{Sec_dis}

As mentioned in Paper I, HD 141943 is only the second \citep[or third including the results for HD 106506 by][]{WaiteIA:2010} young early-G star for which the large-scale magnetic topology has been determined, the other being HD 171488 \citep{MarsdenSC:2006, JeffersSV:2008, JeffersSV:2010}. In total there have been five young early-G stars for which differential rotation measures have been determined. The five stars are HD 141943, HD 171488, HD 106506, R58 \citep{MarsdenSC:2005a, MarsdenSC:2005b} and LQ Lup \citep{DonatiJF:2000}. The stellar parameters for all five stars have been given in Paper I. However, for ease of comparison we have given these in Table~\ref{Tab_gstars} and included the differential rotation results of the five stars.

\begin{table*}
\caption{Comparison of the stellar parameters of the five young early-G stars that have had their surface differential rotation measured using Doppler imaging. Except where noted, the data for HD 141943 come from this work and Paper I, that for HD 106506 is from \citet{WaiteIA:2010} and that for HD 171488 from \citet{StrassmeierKG:2003}, \citet{MarsdenSC:2006}, \citet{JeffersSV:2008} and \citet{JeffersSV:2010}. The data for R58 come from \citet{MarsdenSC:2005a,MarsdenSC:2005b} and the data for LQ Lup from \citet{DonatiJF:2000}. The values for the depth of the convective zone are from \citet{SiessL:2000}. For those stars with multiple measurements of differential rotation a value has been used with error bars large enough to encompass the entire range of the observations.}
\label{Tab_gstars}
\centering
\begin{tabular}{lccccc}
\hline\hline
Parameter & HD 141943 & HD 106506 & HD 171488 & R58 & LQ Lup \\
\hline
(B-V)                                                          & 0.65$^{a}$                                      & 0.605                                                  & 0.62$^{a}$                                    & 0.61$^{b}$                                        & 0.69$^{c}$            \\
Age (Myrs)                                                & $\sim$17                                         & $\la$10                                              & 30 -- 50                                          & 35 $\pm$ 5                                       & 25 $\pm$ 10         \\
Mass (M\subs{\odot})                              & $\sim$1.3                                        & 1.5 $\pm$ 0.1                                   & 1.20 $\pm$ 0.02                           & 1.15 $\pm$ 0.05                             & 1.16 $\pm$ 0.04  \\
Radius (R\subs{\odot})                           & 1.6 $\pm$ 0.1                                 & 2.15 $\pm$ 0.26                              & 1.15 $\pm$ 0.08                           & 1.18$^{+0.17}_{-0.10}$                 & 1.22 $\pm$ 0.12   \\
Inclination (\sups{\circ})                          & 70 $\pm$ 10                                   & 65 $\pm$ 5                                       & 60 $\pm$ 10                                  & 60 $\pm$ 10                                    & 35 $\pm$ 5            \\
Convective zone (R\subs{\star})           & $\sim$0.16 [0.26 R\subs{\odot}] & $\sim$0.22 [0.47 R\subs{\odot}] & $\sim$0.21 [0.24 R\subs{\odot}] & $\sim$0.21 [0.25 R\subs{\odot}] & $\sim$0.26 [0.32 R\subs{\odot}] \\
$\Omega$\subs{eq} (rad d\sups{-1}) Stokes I   & 2.86 $\pm$ 0.02          & 4.54 $\pm$ 0.01 & 4.84 $\pm$ 0.14$^d$ & 11.16 $\pm$ 0.04$^d$ & 20.28 $\pm$ 0.01\\ 
$\Omega$\subs{eq} (rad d\sups{-1}) Stokes V & 2.89 $\pm$ 0.05$^d$ & 4.51 $\pm$ 0.01 & 4.81 $\pm$ 0.10$^d$ & --                                       & --                              \\ 
$d\Omega$ (rad d\sups{-1}) Stokes I                 & 0.24 $\pm$ 0.03          & 0.21 $\pm$ 0.03 & 0.41 $\pm$ 0.16$^d$ & 0.08 $\pm$ 0.07$^d$   & 0.12 $\pm$ 0.02    \\
$d\Omega$ (rad d\sups{-1}) Stokes V                & 0.40 $\pm$ 0.13$^d$ & 0.24 $\pm$ 0.03 & 0.45 $\pm$ 0.06$^d$ & --                                      & --                               \\
\hline
\end{tabular}
\\
$^a$from \citet{CutispotoG:2002}; $^b$from \citet{RandichS:2001}, dereddened value; $^c$from \citet{WichmannR:1997}; $^d$multiple observations.
\end{table*}

\subsection{Surface differential rotation} \label{Sec_ddr}

The differential rotation rate found for HD 14143 is one of the largest yet found using the Doppler imaging method and is similar to that of the other young early-G star HD 171488 \citep{MarsdenSC:2006, JeffersSV:2008, JeffersSV:2010}. It is also similar in level to that of the more mature planet-hosting late-F star Tau Boo \citep{DonatiJF:2008, FaresR:2009} and is in agreement with the findings of high levels of differential rotation on inactive F stars by \citet{ReinersA:2006} using the line-profile method to measure differential rotation. However, the level of differential rotation for HD 141943 and HD 171488 are significantly above that of the other young early-G stars studied using the Doppler imaging method (see Table~\ref{Tab_gstars}). There appears to be little differences between the stars to explain the differences in differential rotation with the possible exception of their convective zone depth.

\begin{figure}
  \centering
  \includegraphics[angle=90, width=\columnwidth]{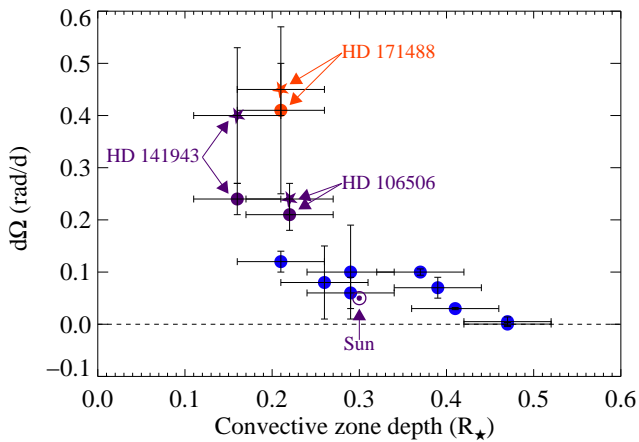}
  \caption{Differential rotation ($d\Omega$) versus convective zone depth (as a function of stellar radius) for young solar-type stars from \citet{BarnesJR:2005} and the new stars in Table~\ref{Tab_gstars}. The convective zone depth has been determined from \citet{SiessL:2000} with an assumed error of $\pm$0.05 R\subs{\star}. Dots show differential rotation measurements using brightness features and stars show differential rotation measures from magnetic features. For stars with multiple measurements a value has been used with error bars large enough to encompass the entire range of the observations.}
  \label{Fig_czddr}
\end{figure}

\citet{BarnesJR:2005} have shown that generally differential rotation increases with stellar effective temperature, however, their dataset does not have any stars with a differential rotation greater than $d\Omega$ $\sim$ 0.2 rad d\sups{-1}. As an extension of the work of \citet{BarnesJR:2005} we have plotted the differential rotation measured using the (Zeeman) Doppler imaging technique for these stars plus the additional stars listed in Table~\ref{Tab_gstars} (HD141943, HD 106506 and HD 171488). Fig.~\ref{Fig_czddr} plots this differential rotation against stellar convective zone depth \citep*[determined from][]{SiessL:2000}.

As can be seen in Fig.~\ref{Fig_czddr} the level of differential rotation does appear to increase slightly with decreasing convection zones depth (with some scatter) until  the convective zone depth reaches $\sim$0.2 R\subs{\star} (early-G stars) and then a dramatic increase in the level of surface differential rotation occurs. Why this should be is still not understood, but it appears that a change occurs in the rotation of the convective zone at this depth. As suggested by \citet{JeffersSV:2008}, for stars with thinner convective zones, such as HD 141943, we may be seeing closer to the base of the convective zone and this could explain the higher levels of differential rotation seen on these stars. 

Results from \citet{DonatiJF:2003b} have shown that for early-K stars the differential rotation measured from magnetic features is higher than that measured from brightness features. They attribute this to the brightness and magnetic features being anchored at different depths in the stellar convective zone and the convective zone having a radially varying differential rotation (unlike the Sun). \citet{JeffersSV:2008} and \citet{JeffersSV:2010} found that there is virtually no change (within errors) in the differential rotation measured from brightness and magnetic features for the early-G star HD 171488 although the errors in the differential rotation measurements are, for the most part, much larger than the errors for the early-K stars. Although their Stokes V differential rotation measurements are for the most part higher than those measured from Stokes I. HD 106506 \citep{WaiteIA:2010} also shows only a small increase in  the differential rotation from magnetic features over that from brightness features, but again the difference is within the errors of the measurements.  

In contrast to this our 2010 results for HD 141943 show a large difference between the differential rotation measured from the brightness features compared to that measured from the magnetic features, with the magnetic differential rotation being significantly higher than that from the brightness features. One possible reason for this difference could be the different latitude distributions of the brightness and magnetic features on HD 141943. In 2010 the spot features of HD 141943 are concentrated in a minor polar spot and some lower latitude features, while the magnetic features are more evenly distributed over the entire hemisphere (see Fig.~5 in Paper I). Such a difference in distribution may possibly lead to biases in the determination of the differential rotation, as the differential rotation is determined over a smaller (or larger) latitude range. Both HD 106506 and HD 171488 have larger polar spots and less low-latitude spot features than HD 141943. Thus, the two stars with more dominant polar spots would be expected to more affected by any bias in the latitude distribution between brightness and magnetic features, but neither star shows any significant disparity in differential rotation. The reason why HD 141943 alone shows a disparity in the differential rotation measurements between brightness and magnetic features remains unknown. A larger sample size is required to determine if HD 141943 is just a unique case.

Unlike the results from early-K stars \citep{DonatiJF:2003b, JeffersSV:2007} the early-G star HD 171488 shows no evidence of temporal evolution in its level of differential rotation \citep{MarsdenSC:2006, JeffersSV:2008, JeffersSV:2010} although again the errors are larger than the level of variation seen on early-K stars. Our magnetic results from 2007 and 2010 for HD 141943 show a slight difference (see Section~\ref{Sec_sdr}) but with only a $\sim$2$\sigma$ change this is only rather tentative evidence for temporal evolution in the star's differential rotation. The temporal variability seen on early-K stars is seen as evidence of a dynamo that periodically converts magnetic energy into kinetic energy, and vise-versa, in the stellar convective zone. If such a mechanism is occurring in early-G stars (with thinner convective zones) it does not appear to have a large impact on the level of their differential rotation, although the errors in our differential rotation measures are too great to rule out a level of variability similar to that seen on early-K stars. 

Due to the rapid winding of magnetic field lines most likely to occur on stars with high levels of differential rotation, it is possible that such stars could have significantly shorter magnetic cycles than the 22-year solar magnetic cycle. Indeed the late-F star Tau Boo does in fact appear to have a very short magnetic cycle \citep[of $\sim$2 years,][]{FaresR:2009}, but this star is also the host to a ``Hot Jupiter'' \citep{ButlerRP:1997} which may also be affecting its magnetic cycle length. Both the early-G stars HD 141943 and HD 171488 have similar levels of differential rotation to that of Tau Boo and do not show any evidence of a magnetic polarity reversal over $\sim$3 years of observations \citep{JeffersSV:2010, MarsdenSC:2010}. So the role of differential rotation in the magnetic cycles (or if they even have cycles) of these young stars is still unknown.

\subsection{Coronal magnetic field} \label{Sec_cormag}

Fig.~\ref{cor_struct} shows the structure of the coronal magnetic field and also of the X-ray emission for the 2007, 2009 and 2010 magnetic maps (see Paper I). Open field lines that would be X-ray dark and carrying the stellar wind, are shown blue, while the X-ray bright closed field lines are shown white. On the largest scales, the dipole component of the field dominates and the field structure for the three years is similar. The main change on the large scale is in the tilt of the dipole component of the field. Between 2007 and 2009, the dipole axis shifted from a latitude of 68 degrees to down to 52 degrees and then back to 72 in 2010. On smaller scales, the field structure is very different between the three years and this is reflected in the X-ray images. As noted in Paper I the poor quality of the 2009 dataset may play some role in these results.

\begin{figure*}
 \begin{center}
  \includegraphics[width=8.5cm]{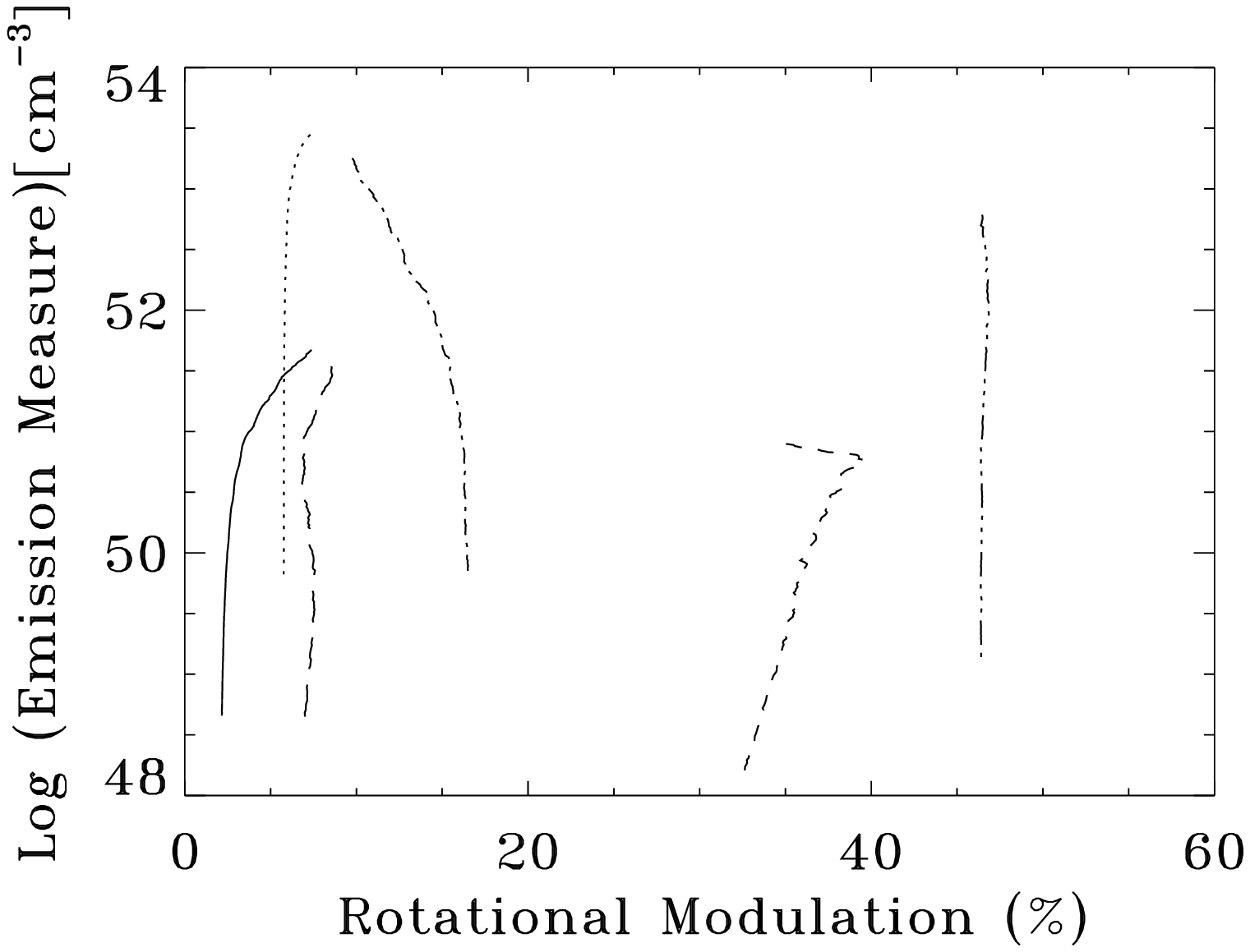}
  \includegraphics[width=8.5cm]{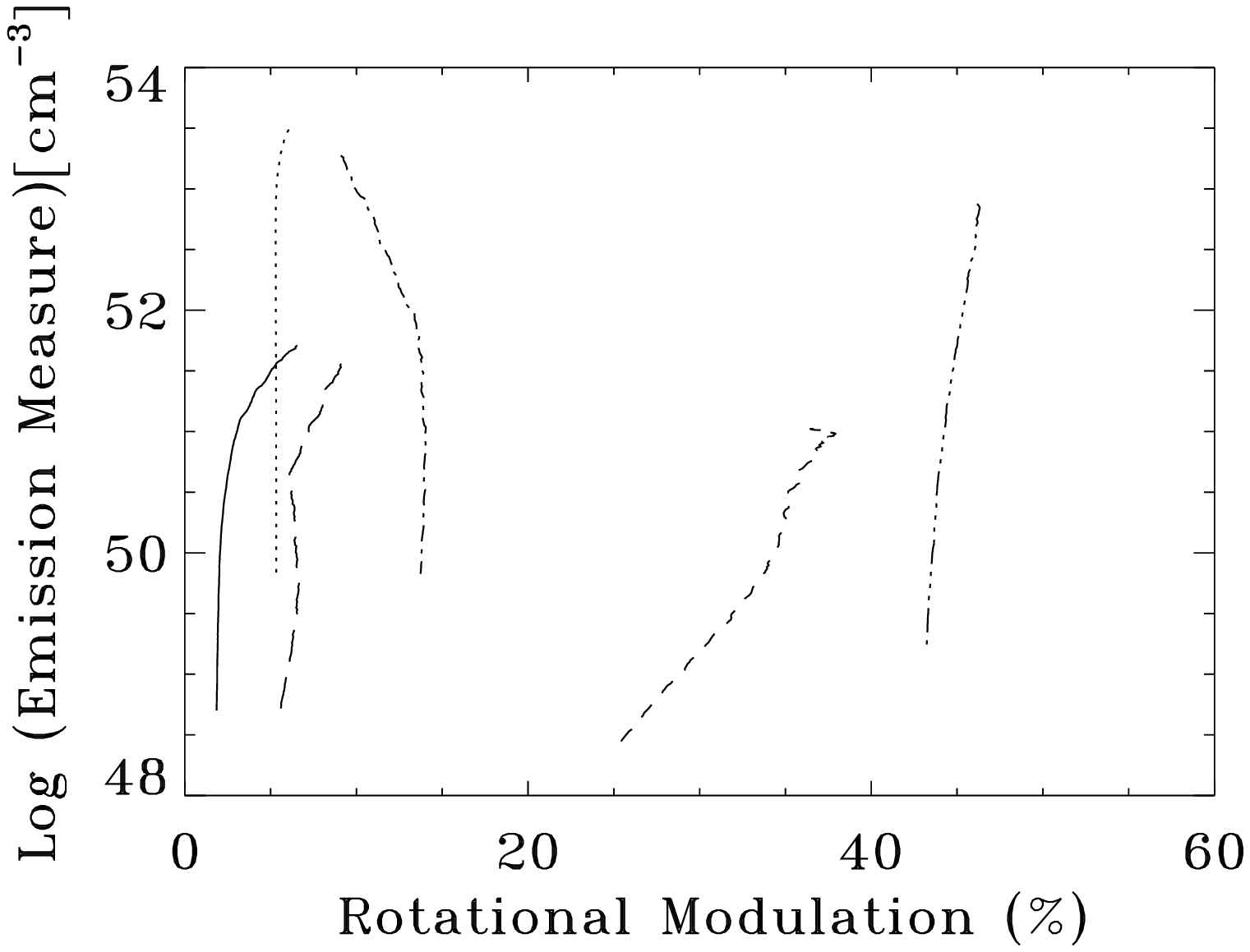}
  \includegraphics[width=8.5cm]{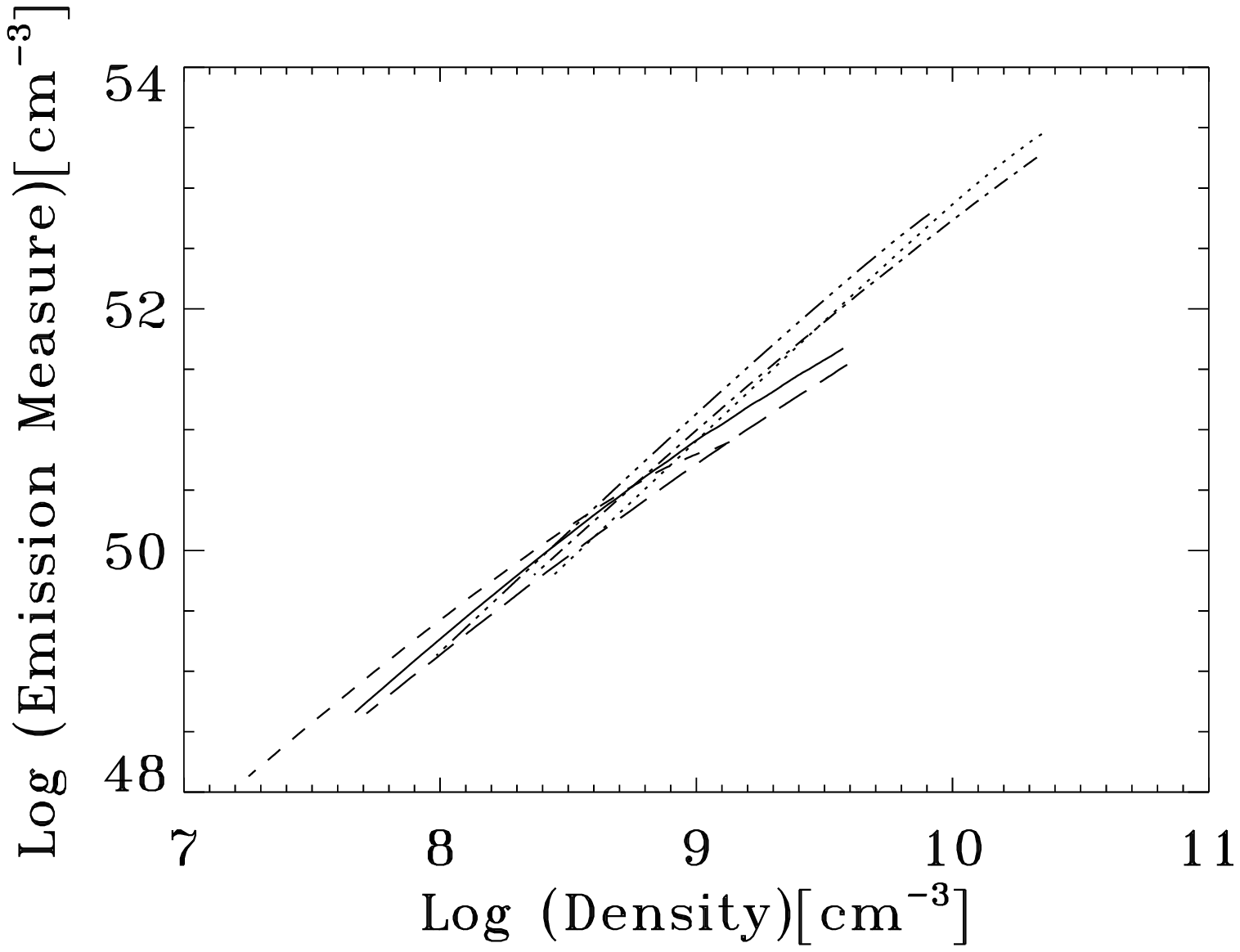}
  \includegraphics[width=8.5cm]{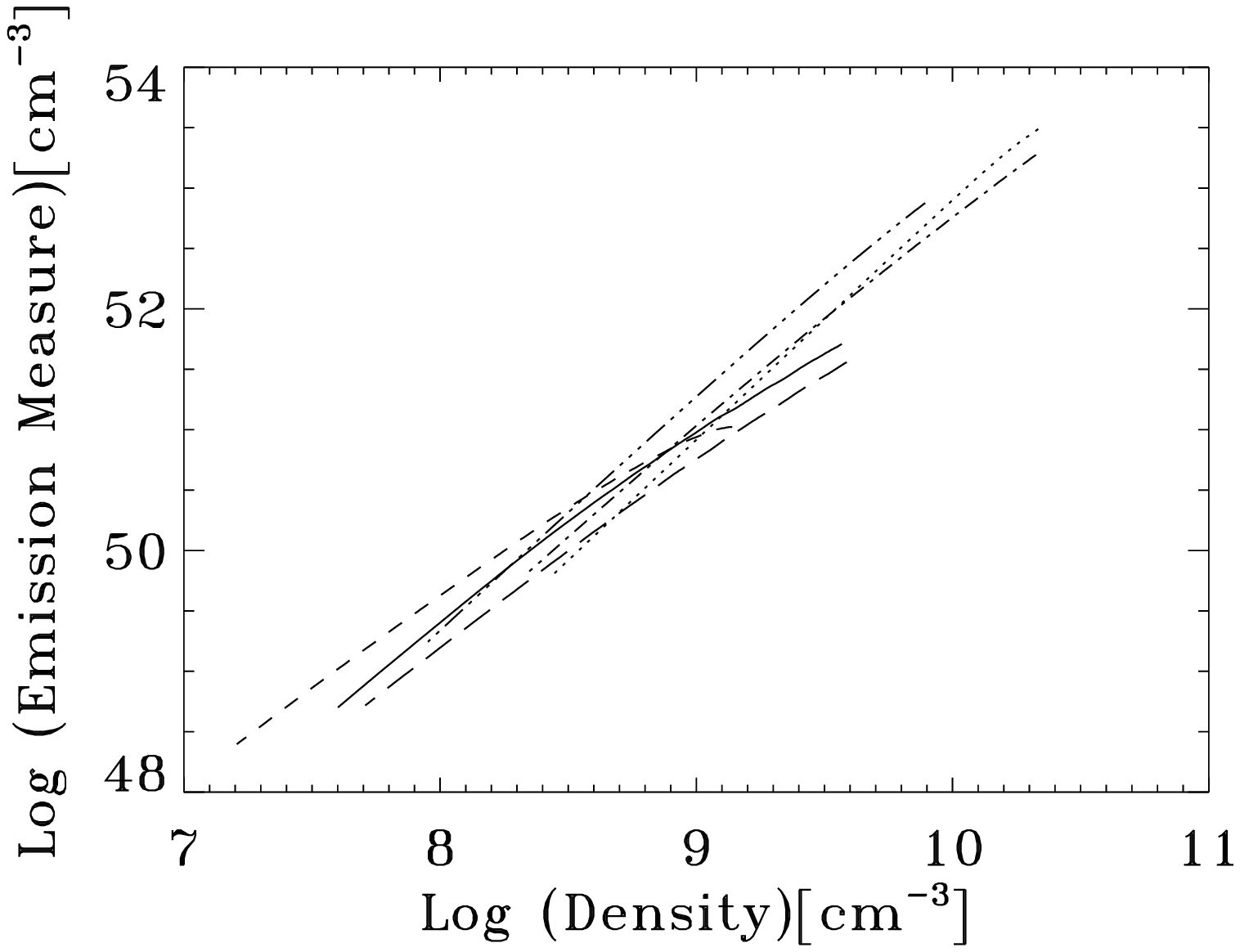}
   \caption{Emission measure for a range of models with varying rotational modulation of the emission measure (top) and also varying densities (bottom). The left hand column shows results for the smaller corona (extending to 2.5 R\subs{\star}) while the right hand column shows results for the larger corona, (extending to 4.8 R\subs{\star}). The linestyles denote different temperatures and data from different years. Thus we have: 2007, T = 2$\times$10\sups{6} K (dotted); 2007, T = 2$\times$10\sups{7} K (solid); 2009, T = 2$\times$10\sups{6} K (triple-dot dashed); 2009, T = 2$\times$10\sups{7} K (short dashed), 2010, T = 2$\times$10\sups{6} K (dot dashed) and 2010, T = 2$\times$10\sups{7} (long dashed).}
  \label{cor_mod}
  \end{center}
\end{figure*}

Fig.~\ref{cor_mod} shows the magnitude and rotational modulation of the emission measure, and also the emission-measure weighted density:
\begin{equation}
\bar{n}_e =\frac{ \int n_e^3 dV}{ \int n_e^2 dV} .
\end{equation}
These are calculated for a range of values of the base pressure, since without simultaneous X-ray observations we are unable to determine the value of base pressure. We also show in the left column results for a model with a small corona (extending to 2.5 R\subs{\star}) and in the right column results for a model with a larger corona (extending to 4.8 R\subs{\star}). It is immediately apparent that the small changes introduced by uncertainty in the size of the corona are small compared to the changes between years. We expect the corona to be populated with loops at a range of temperatures, but in the absence of any data that might determine the nature of the differential emission measure, we take the simplest approach. We therefore show results for a corona at a uniform temperature of 2$\times$10\sups{6} K and for comparison, one at a uniform temperature of 2$\times$10\sups{7} K.

As might be expected, at the lower temperatures, the densities and emission measures reach higher values. The range of values is however similar for all years and is typical for active stars. The largest difference between the three years is in the arrangement of the smaller scale loops and this results in a significant increase in the X-ray rotational modulation in 2009.

\section{Conclusions} \label{Sec_con}

We have extrapolated the coronal magnetic field of HD 141943 from surface magnetograms acquired at 3 epochs. These show that the large-scale field structure of the corona is dominated by a dipole component with the axis of the dipole shifting between the 3 epochs. The small scale structure shows an increase in the modelled rotational X-ray modulation in 2009, compared to the other epochs (2007 \& 2010).

The surface differential rotation of HD 141943 has been measured at 2 epochs (2007 \& 2010) from both the surface brightness features (2010) and the surface magnetic features (2007 \& 2010). The differential rotation measured from the magnetic features is one of the highest values of differential rotation measured on a young solar-type star and is similar in level to the other young early-G star HD 171488, but higher than the more swollen star HD 106506. We thus conclude that the depth of the stellar convective zone plays a strong role in the level of surface differential rotation seen on solar-type stars, with a large increase in differential rotation seen for star's with convective zone depths shallower than $\sim$0.2 R\subs{\star}.

The 2010 dataset for HD 141943 shows a large increase in the level of differential rotation measured from magnetic features to that measured from brightness features. This is similar to that seen on early-K stars but with a much greater difference and is in contrast to the results from other early-G stars which show little or no difference between the differential rotation measured from brightness and magnetic features. Our results only find tentative evidence for temporal evolution in the differential rotation of HD 141943. These results when combined with those from the early-G star HD 171488 (which shows no evidence of temporal evolution in differential rotation) imply that early-G stars do not undergo large-scale evolution in their differential rotation. However, the errors in our measurements are too large to rule out small scale evolution in differential rotation similar to that seen on early-K stars.

HD 141943 and stars of similar spectral type warrant further observations to determine what effect a shallow convective zone has on the differential rotation levels of such stars and indeed if they do show temporal evolution of their differential rotation as seen on early-K stars.

\section*{Acknowledgments}

The observations in this paper were obtained with the Anglo-Australian telescope. We would like to thank the technical staff of the Anglo-Australian Observatory (now the Australian Astronomical Observatory) for their, as usual, excellent assistance during these observations. We would also like to thank the anonymous referee who helped improve this paper. This project is supported by the Commonwealth of Australia under the International Science Linkages program.



\label{lastpage}


\begin{thebibliography}{}
\bibitem[\protect\citeauthoryear{Altschuler \& Newkirk, Jr.}{1969}]{AltschulerMD:1969} Altschuler M.D., Newkirk, Jr. G., 1969, Solar Phys., 9, 131
\bibitem[\protect\citeauthoryear{Barnes et al.}{2005}]{BarnesJR:2005} Barnes J. R., Collier Cameron A., Donati J.-F., James D. J., Marsden S. C., Petit. P., 2005, MNRAS, 357, L1
\bibitem[\protect\citeauthoryear{Butler et al.}{1997}]{ButlerRP:1997} Butler R. P., Marcy G. W., Williams E., Hauser H., Shirts P., 1997, ApJ, 474, L115
\bibitem[\protect\citeauthoryear{Collier Cameron \& Robinson}{1989}]{CameronAC:1989} Collier Cameron A., Robinson R.D., 1989, MNRAS, 236, 57
\bibitem[\protect\citeauthoryear{Collier Cameron, Donati \& Semel}{Collier Cameron et al.}{2002}]{CameronAC:2002} Collier Cameron A., Donati J.-F., Semel M., 2002, MNRAS, 330, 699
\bibitem[\protect\citeauthoryear{Cutispoto et al.}{2002}]{CutispotoG:2002} Cutispoto G., Pastori L., Pasquini L., de Medeiros J. R., Tagliaferri G., Anderson J., 2002, A\&A, 384, 491
\bibitem[\protect\citeauthoryear{Donati \& Collier Cameron}{1997}]{DonatiJF:1997a} Donati J.-F., Collier Cameron A., 1997, MNRAS, 291, 1
\bibitem[\protect\citeauthoryear{Donati et al.}{1997}]{DonatiJF:1997b} Donati J.-F., Semel M., Carter B. D., Rees D. E., Cameron A. C., 1997, MNRAS, 291, 658
\bibitem[\protect\citeauthoryear{Donati}{1999}]{DonatiJF:1999b} Donati J.-F., 1999, MNRAS, 302, 457
\bibitem[\protect\citeauthoryear{Donati et al.}{1999}]{DonatiJF:1999a} Donati J.-F., Collier Cameron A., Hussain G., Semel M., 1999, MNRAS, 302, 437
\bibitem[\protect\citeauthoryear{Donati et al.}{2000}]{DonatiJF:2000} Donati J.-F., Mengel M., Carter B. D., Marsden S., Collier Cameron A., Wichmann R., 2000, MNRAS, 316, 699
\bibitem[\protect\citeauthoryear{Donati et al.}{2003a}]{DonatiJF:2003a} Donati J.-F., Collier Cameron A., Semel M., et al., 2003a, MNRAS, 345, 1145
\bibitem[\protect\citeauthoryear{Donati, Collier Cameron \& Petit}{Donati et al.}{2003b}]{DonatiJF:2003b} Donati J.-F., Collier Cameron A., Petit P., 2003b, MNRAS, 345, 1187
\bibitem[\protect\citeauthoryear{Donati et al.}{2008}]{DonatiJF:2008} Donati J.-F., Moutou C., Far\`{e}s R., et al. 2008, MNRAS, 385, 1179
\bibitem[\protect\citeauthoryear{Dunstone et al.}{2008}]{DunstoneNJ:2008} Dunstone H. J., Hussain G. A. J., Collier Cameron A., Marsden S. C., Jardine M., Stempels H. C., Ramirez Vlez J. C., Donati J.-F., 2008, MNRAS, 387, 481
\bibitem[\protect\citeauthoryear{Fares et al.}{2009}]{FaresR:2009} Fares R., Donati J.-F., Moutou C., et al., 2009, MNRAS, 398, 1383
\bibitem[\protect\citeauthoryear{Jardine, Collier Cameron \& Donati}{Jardine et al.}{2002a}]{JardineM:2002a} Jardine M., Collier Cameron A., Donati J.-F., 2002a, MNRAS, 333, 339
\bibitem[\protect\citeauthoryear{Jardine et al.}{2002b}]{JardineM:2002b} Jardine M., Wood K., Collier Cameron A., Donati J.-F., Mackay D. H., 2002b, MNRAS, 336, 1364
\bibitem[\protect\citeauthoryear{Jeffers, Donati \& Collier Cameron}{Jeffers et al.}{2007}]{JeffersSV:2007} Jeffers S. V., Donati J.-F., Collier Cameron A., 2007, MNRAS, 375, 567
\bibitem[\protect\citeauthoryear{Jeffers \& Donati}{2008}]{JeffersSV:2008} Jeffers S. V., Donati J.-F., 2008, MNRAS, 390, 635
\bibitem[\protect\citeauthoryear{Jeffers et al.}{2010}]{JeffersSV:2010} Jeffers S. V., Donati J.-F., Alecian E., Marsden S. C., 2010, MNRAS, accepted
\bibitem[\protect\citeauthoryear{Kitchatinov \& R\"{u}diger}{1999}]{KitchatinovLL:1999} Kitchatinov L. L., R\"{u}diger G., 1999, A\&A, 344, 911
\bibitem[\protect\citeauthoryear{K\"{u}ker \& R\"{u}diger}{2005}]{KukerM:2005} K\"{u}ker M., R\"{u}diger G., 2005, AN, 326, 265
\bibitem[\protect\citeauthoryear{Marsden et al.}{2005a}]{MarsdenSC:2005a} Marsden S. C., Carter B. D., Donati J.-F., 2005a, in Favata F., Hussain G. A. J., Battrick B., eds. Proceedings of the 13\sups{th} Cambridge Workshop on Cool Stars, Stellar Systems and the Sun, ESA Special Publications, Vol. 560, ESA, Noordwijk, The Netherlands, p. 799
\bibitem[\protect\citeauthoryear{Marsden et al.}{2005b}]{MarsdenSC:2005b} Marsden S. C., Waite I. A., Carter B. D., Donati J.-F., 2005b, MNRAS, 359, 711
\bibitem[\protect\citeauthoryear{Marsden et al.}{2006}]{MarsdenSC:2006} Marsden S. C., Donati J.-F., Semel M., Petit P., Carter B. D., 2006, MNRAS, 370, 468
\bibitem[\protect\citeauthoryear{Marsden et al.}{2010}]{MarsdenSC:2010} Marsden S. C., Jardine M. M., Ram\'{i}rez V'{e}lez J. C., et al., 2010, MNRAS, accepted (Paper I)
\bibitem[\protect\citeauthoryear{Petit, Donati \& Collier Cameron}{Petit et al.}{2002}]{PetitP:2002} Petit P., Donati J.-F., Collier Cameron A., 2002, MNRAS, 334, 374
\bibitem[\protect\citeauthoryear{Randich}{2001}]{RandichS:2001} Randich S., 2001, A\&A, 377, 512
\bibitem[\protect\citeauthoryear{Reiners \& Schmitt}{2002}]{ReinersA:2002} Reiners A., Schmitt J. H. M. M., 2002, A\&A, 384, 155
\bibitem[\protect\citeauthoryear{Reiners \& Schmitt}{2003}]{ReinersA:2003} Reiners A., Schmitt J. H. M. M., 2003, A\&A, 398, 647
\bibitem[\protect\citeauthoryear{Reiners}{2006}]{ReinersA:2006} Reiners A., 2006, A\&A, 446, 267
\bibitem[\protect\citeauthoryear{Semel, Donati \& Rees}{Semel et al.}{1993}]{SemelM:1993} Semel M., Donati J.-F., Rees D. E., 1993, A\&A, 278, 231
\bibitem[\protect\citeauthoryear{Siess, Dufour \& Forestini}{Siess et al.}{2000}]{SiessL:2000} Siess L., Dufour E., Forestini M., 2000, A\&A, 358, 593
\bibitem[\protect\citeauthoryear{Soderblom et al.}{1993}]{SoderblomDR:1993} Soderblom D. R., Stauffer J. R., Hudon J. D., Jones B. F., 1993, ApJS, 85, 315
\bibitem[\protect\citeauthoryear{Strassmeier et al.}{2003}]{StrassmeierKG:2003} Strassmeier K. G., Pichler T., Weber M., Granzer T., 2003, A\&A, 411, 595
\bibitem[\protect\citeauthoryear{van Ballegooijen, Cartledge \& Priest}{van Ballegooijen et al.}{1998}]{vanBallegooijenA:1998} van Ballegooijen A., Cartledge N., Priest E., 1998, ApJ, 501, 866
\bibitem[\protect\citeauthoryear{Waite et al.}{2010}]{WaiteIA:2010} Waite I. A., Marsden S.C., Carter B.D., et al., 2010, MNRAS, accepted
\bibitem[\protect\citeauthoryear{Wichmann et al.}{1997}]{WichmannR:1997} Wichmann R., Krautter J., Covino E., Alcal\'{a} J. M., Neuh\"{a}user R., Schmitt J. H. M. M., 1997, A\&A, 320, 185
\end{thebibliography}
\end{document}